\documentclass[prb,twocolumn,showpacs]{revtex4}
\usepackage[latin1]{inputenc}
\usepackage{graphicx}
\usepackage{amssymb}
\usepackage{amsmath}
\usepackage{xspace}
\usepackage{dcolumn}
\usepackage{bm}

\newcommand{\ie}[0]{i.e.\@\xspace}
\newcommand{\eg}[0]{e.g.\@\xspace}
\newcommand{\etal}[0]{{\it et al.}\@\xspace}
\newcommand{\Z}[0]{\mathcal{Z}}
\newcommand{\D}[0]{\mathcal{D}}
\newcommand{\omb}[0]{\bar{\omega}}

\newcommand{\om}[0]{\omega}
\newcommand{\las}[0]{\langle}
\newcommand{\ras}[0]{\rangle}
\newcommand{\dtau}{\Delta\tau}
\newcommand{\rmi}{\mathrm{i}}
\newcommand{\op}{\hat{p}}
\newcommand{\ox}{\hat{x}}
\newcommand{\on}{\hat{n}}
\newcommand{\wb}{w_\text{b}}
\newcommand{\wf}{w_\text{f}}
\newcommand{\UP}{\uparrow}
\newcommand{\DO}{\downarrow}
\newcommand{\sign}{\las\text{sign}\ras}
\newcommand{\rD}{\text{D}}
\newcommand{\Ek}{\bar{E}_\text{k}}
\newcommand{\lc}{\lambda_\text{C}}
\newcommand{\Ep}{E_\text{P}}

\begin{document}


\title{Holstein polaron in two and three dimensions by quantum Monte Carlo}

\author{Martin Hohenadler}\email{hohenadler@itp.tu-graz.ac.at}
\author{Hans Gerd Evertz}
\author{Wolfgang \surname{von der Linden}}
\affiliation{%
  Institute for Theoretical and Computational Physics, Graz University of
  Technology, Petersgasse 16, A-8010 Graz, Austria}

\begin{abstract}
  A recently developed quantum Monte Carlo approach to the Holstein model
  with one electron [PRB {\bf 69}, 024301 (2004)] is extended to two and
  three dimensional lattices.  A moderate sign problem occurs, which is found
  to diminish with increasing system size in all dimensions, and not to
  affect simulations significantly. We present an extensive study of the
  influence of temperature, system size, dimensionality and model parameters
  on the small-polaron cross over. Results are extrapolated to remove the
  error due to the Trotter discretization, which significantly improves the
  accuracy. Comparison with existing work and other quantum Monte Carlo
  methods is made. The method can be extended to the many-electron case.
\end{abstract}


\maketitle

\section{\label{sec:introduction}Introduction}

In recent years, a lot of experimental evidence has been given for the
importance of electron-phonon interaction in strongly correlated systems such
as the cuprates\cite{BYMdLBi92} or the manganites.\cite{David_AiP} Although
considerable theoretical progress has been made in understanding and
describing many details of the physics of these compounds, the quantum nature
of the phonons has often been neglected in actual
calculations.\cite{MiMuSh96II,DaYuMobook} However, a quantum-mechanical
treatment of the lattice degrees of freedom has been shown to give
significantly different results in certain parameter regimes.\cite{David_AiP}

Due to the complexity of the models for the abovementioned classes of
materials, numerical methods have been used extensively. One very powerful
approach is the quantum Monte Carlo (QMC) method, which allows for
simulations on relatively large lattices and gives quasiexact results (\ie,
exact apart from statistical errors which can, in principle, be made
arbitrarily small) also at finite temperature. The latter point represents a
major advantage over, \eg, the density matrix renormalization group (DMRG) or
variational methods---which are restricted to the calculation of ground-state
properties---since fascinating phenomena such as high-temperature
superconductivity and colossal magnetoresistance can be investigated.  Exact
diagonalization (ED) can be performed at finite temperatures (see, \eg,
Ref.~\onlinecite{AiDaEvvdL02}), but for the case of electron-phonon systems
such calculations would be restricted to very small systems and unrealistic
parameters. Consequently, all existing work based on ED is for zero
temperature only. Additionally, for coupled electron-phonon systems, the
infinite-dimensional Hilbert space associated with the boson degrees of
freedom represents a substantial difficulty for ED and DMRG, in contrast to
QMC. Nevertheless, QMC methods are often limited by (1) the minus-sign
problem, which restricts simulations to high temperatures and/or small
systems, (2) the fact that the calculation of dynamical properties, such as
the one-electron Green function, requires analytic continuation to the
real-time axis which is an ill-posed problem, and (3) by strong
autocorrelations and large statistical errors.

In a recent paper,\cite{HoEvvdL03} from here on referred to as I, we have
proposed a new QMC approach to the Holstein model, which is based on the
canonical Lang-Firsov transformation\cite{LangFirsov} and a principal
component representation of the phonon degrees of freedom. The resulting
algorithm has been shown to completely avoid the problem of autocorrelations,
strongly reduce statistical errors and to allow for very efficient
simulations.\cite{HoEvvdL03} The work here and in I is restricted to the case
of a single electron which is often called the ``{\it Holstein polaron
  problem}''.  Obviously, in connection with materials such as the cuprates
or manganites, we are interested in studying the many-electron system.
However, the one-electron case represents an important first step towards an
understanding of more general situations. Moreover, due to the coupling to
the phonons, even the model with one electron constitutes a complex many-body
problem which has received a lot of attention in the past. An important
feature of our approach is the fact that it can be generalized to the more
realistic many-electron case encountered in transition metal oxides.

In this paper, we extend the QMC method proposed in I to square (cubic)
lattices in two (three) dimensions. Although the one-dimensional case
considered in I is of great theoretical interest, real strongly correlated
materials usually require two or three dimensional models. Consequently,
before turning to the more challenging many-electron case, it is crucial to
examine the influence of dimensionality on our algorithm. To this end, we
investigate the dependence of the minus-sign problem, which has been found to
appear in simulations of the Lang-Firsov transformed model, on the
dimensionality D of the lattice, as well as on the parameters of the model.
We then use our method to study in detail the well-known small-polaron cross
over, which occurs with increasing electron-phonon coupling strength, and its
dependence on phonon frequency, temperature, system size and dimensionality.
We remove the Trotter error, which results from the discretization of
imaginary time, by extrapolating our results to the limit of continuous time,
thereby significantly improving the accuracy of the data.

The paper is organized as follows. In Sec.~\ref{sec:holstein} we review the
physics of the Holstein model with one electron, and in Sec.~\ref{sec:QMC} we
briefly discuss the QMC method introduced in I. Results are presented in
Sec.~\ref{sec:results} and Sec.~\ref{sec:summary} contains our conclusions.

\section{\label{sec:holstein}The Holstein model}

The Holstein Hamiltonian\cite{Ho59a} with dimensionless phonon variables
reads\cite{HoEvvdL03}
\begin{eqnarray}\label{eq:holstein}\nonumber
  H
  &=&
  -t\sum_{\las ij\ras} c^\dag_i c^{\phantom{\dag}}_j
  +\frac{\om}{2}\sum_i\left( \op_i^2 + \ox_i^2 \right)
  -\alpha\sum_i \on_i \ox_i
  \\
  &=&
  K + P + I\,,
\end{eqnarray}
where we have introduced the abbreviations $K$, $P$ and $I$ for the kinetic,
potential and interaction terms, respectively. Spin indices have been
suppressed in the notation, owing to the spin symmetry of the one-electron
problem. In Eq.~(\ref{eq:holstein}), $c^\dag_i$ ($c_i$) creates (annihilates)
a spinless electron at lattice site $i$, $\ox_i$ and $\op_i$ denote the
displacement and momentum operator of a harmonic oscillator at site $i$, and
$\on_{i}=c^\dag_i c^{\phantom{\dag}}_i$. The coupling term $I$ describes the
local interaction of the single electron considered here with dispersionless
Einstein phonons. In the first term, the symbol $\las ij\ras$ denotes a
summation over all hopping processes $i\rightarrow j$ and $j\rightarrow i$
between neighboring lattice sites $i,j$.  The parameters of the model are the
hopping integral $t$, the phonon energy $\om$ ($\hbar=1$), and the
electron-phonon coupling constant $\alpha$. As in I, we define the
dimensionless coupling constant $\lambda = \alpha^2/(\om W)$, where $W=4t\rD$
is the bare bandwidth in D dimensions. We shall also use the dimensionless
phonon frequency $\omb=\om/t$, also called the ``{\it adiabatic ratio}'', and
express all energies in units of $t$. Consequently, the model can be
described by two independent parameters, $\omb$ and $\lambda$. While the bulk
of results in this paper has been calculated assuming periodic boundary
conditions in real space, we shall also discuss the effect of the boundary
conditions on the sign problem in Sec.~\ref{sec:sign-problem}.

Since the literature on the Holstein polaron problem is vast, we restrict our
discussion to work in more than one dimension. A brief review of the
one-dimensional model can be found in I. Moreover, we will focus on recent
progress in the field, and on numerical methods. A comprehensive overview of
earlier analytical work can be found, \eg, in the books of Alexandrov and
Mott\cite{AlMo95} and Mahan.\cite{Ma90} The Holstein polaron in $\rD>1$ has
been studied using a large variety of numerical techniques. In contrast to
many perturbative approaches,\cite{AlMo95,Ma90} the latter can also
accurately describe the physically most interesting regime of small but
finite phonon frequency ($0<\omb<1$), and intermediate electron-phonon
coupling ($\lambda\approx1$). Much information about the Holstein polaron has
been obtained using QMC. De Raedt and Lagendijk\cite{dRLa82,dRLa83,dRLa84}
applied Feynman's path-integral technique to the lattice problem of
Eq.~(\ref{eq:holstein}).  In this approach, the phonon degrees of freedom are
integrated out analytically, and one is left with a single-particle system
with a retarded self-interaction, which can be simulated using the Monte
Carlo method. The only approximation consists of discretizing the imaginary
time using the Suzuki-Trotter decomposition.\cite{dRLa83} As the phonons have
been completely eliminated from the problem, very efficient simulations on
large lattices even in higher dimensions are possible. Moreover, the method
is not restricted to the Holstein Hamiltonian~(\ref{eq:holstein}). It can be
extended to include long-range electron-phonon coupling as well as dispersive
phonons.\cite{dRLa84} Kornilovitch\cite{Ko97} later employed the same method
to study the whole range of values of the adiabatic ratio $\omb$, and
extrapolated the results to remove the error due to the Trotter
approximation. He also pointed out the connection of the algorithm to
world-line methods.\cite{Ko97} Extending this work, Kornilovitch developed a
QMC approach which is formulated in continuous imaginary time, and which is
capable of directly measuring the polaron band dispersion $E(\bm{k})$ and the
density of states in one to three dimensions, by sampling world lines with
open boundary conditions in imaginary time.\cite{Ko98,Ko99} Although the
method gives very accurate results and can also be applied to more general
models with, \eg, long-range interaction, it is limited to the regime of
intermediate and strong electron-phonon coupling as well as $\omb\gtrsim 1$
by a minus-sign problem.\cite{Ko98,Ko99} Consequently, it is not as
universally applicable as the algorithm of de Raedt and
Lagendijk.\cite{dRLa82,dRLa83,dRLa84,Ko97} As the above discussion reveals,
the QMC methods of Refs.~\onlinecite{dRLa82,dRLa83,dRLa84,Ko97,Ko98,Ko99} are
very well suited to study the one-electron problem. Moreover, de Raedt and
Lagendijk also extended their approach to the bipolaron problem of two
electrons with opposite spins.\cite{deRaLa86} However, as pointed out by
Kornilovitch,\cite{Ko97} these are all world-line methods. Consequently,
despite the possibility of integrating out the phonon degrees of freedom even
in the many-electron case, they face a serious sign problem in more than one
dimension, for two or more fermions of the same spin, and can therefore not
be used to study many-particle systems. In the context of superconductivity,
the Holstein and the Holstein-Hubbard model with many electrons have been
investigated\cite{BlScSu81,ScSu81,LeSu90,LeSu91,NiGuScFo93} using the
grand-canonical determinant method of Blankenbecler \etal\cite{BlScSu81} (see
also I). However, as discussed in I, due to strong autocorrelations, these
simulations were restricted to rather large phonon frequencies $\omb\geq1$,
while, \eg, the cuprates and the manganites fall into the adiabatic regime
$\omb\ll1$. This is exactly the point where our new approach enters. As we
shall see below, it completely avoids the problem of autocorrelations.

Apart from QMC, several other methods have been successfully applied to the
Holstein polaron. This includes ED in combination with a truncation of the
phonon Hilbert space,\cite{FeRoWe95,WeRoFe96,WeFe97} finite-cluster strong
coupling perturbation theory (SCPT),\cite{Stephan} cluster perturbation
theory (CPT),\cite{HoAivdL03} DMRG,\cite{JeWh98} and a variational
diagonalization method.\cite{KuTrBo02} As pointed out before, standard ED is
restricted to rather small numbers of lattice sites and/or phonon states,
while DMRG\cite{JeWh98} and the variational method of
Ref.~\onlinecite{KuTrBo02} are applicable over a wider range of values of
phonon frequency and electron-phonon coupling, and are much less influenced
by finite-size effects. The same is true for SCPT and CPT, which exactly
diagonalize small clusters---for which enough phonon states can be included
in the calculation---and extrapolate the results to the thermodynamic limit
by treating the rest of the system as a perturbation.\cite{Stephan,HoAivdL03}
Nevertheless, as mentioned before, all this work was limited to zero
temperature. Finally, the Holstein polaron has been investigated recently
using weak- and strong-coupling perturbation
theory,\cite{RoBrLi99,RoBrLi99III} and a variational wave function which is a
superposition of Bloch states for the weak- and the strong-coupling
regime.\cite{CadFIa99}

From all this work, many properties of the Holstein polaron are well
understood. Similar to one dimension,\cite{HoEvvdL03} there is a cross over
from a quasiparticle with slightly increased effective mass to a heavy small
polaron as the electron-phonon coupling strength increases. As pointed out by
Fehske \etal\cite{FeRoWe95} and Capone \etal,\cite{CaStGr97} the two
conditions for the existence of a small polaron are $\lambda>1$ and
$\lambda\rD/\omb>1$. We will first discuss the weak-coupling state.
Concerning its nature in higher dimensions, there exist two different views.
From calculations based on the adiabatic approximation, \ie, taking the limit
$\omb\rightarrow0$, one expects a qualitatively different behavior in one
dimension compared to $\rD>1$, also for $\omb>0$. Wellein
\etal\cite{WeRoFe96} distinguish between the adiabatic ($\omb<1$) and the
nonadiabatic ($\omb>1$) regime. In the adiabatic case, and for $\rD>1$, the
electron is expected to remain quasifree with an almost unchanged effective
mass and kinetic energy for $\lambda<1$, corresponding to a quasiparticle
with infinite radius. This contrasts strongly with the one-dimensional case,
in which the electron is always self-trapped by the surrounding lattice
distortion and forms a polaron with finite radius for any $\lambda>0$. On the
other hand, in the nonadiabatic regime $\omb>1$, the behavior is very similar
in all dimensions, and a very gradual decrease of the kinetic energy (or
increase in effective mass) is observed as the coupling strength increases.
Due to the large energy of the phonon excitations, only the zero-phonon state
contributes significantly. Moreover, the phonons are fast and react almost
instantly to the motion of the electron. Consequently, a lattice distortion
only persists in the immediate vicinity of the electron, and this rather
small quasiparticle is sometimes called a ``{\it nonadiabatic Lang-Firsov
  polaron}''.\cite{WeRoFe96} Romero \etal\cite{RoBrLi99III} take on a
slightly different viewpoint. They argue that the large polaron state is
essentially the same in any dimension, and that the only effect of increasing
the dimension of the system is given by the observed sharpening of the
transition to a small polaron in $\rD>1$ compared to 1D. For $\lambda$ larger
than a critical value $\lc$, determined by the aforementioned conditions
$\lambda>1$ and $\lambda\rD/\omb>1$, where the cross over occurs, both views
agree on the existence of a a small-sized so-called ``{\it Holstein
  polaron}''.\cite{WeRoFe96} The latter is a heavy quasiparticle with a
strongly reduced mobility.  The cross over at $\lc$ is very sharp, especially
for $\omb\ll1$, but it does not represent a real phase
transition.\cite{Loe88} Even though the electron is trapped in the potential
well originating from the response of the lattice to its motion, the ground
state is still Bloch-like. For simplicity, in the remainder of this paper, we
shall always refer to the weak-coupling state as a large polaron, either with
finite or infinite radius, depending on which of the abovementioned
viewpoints one holds. In our opinion, a definite decision about which of the
two alternatives is correct cannot be made using QMC, which is restricted to
finite clusters and, more importantly, finite temperatures. As a consequence,
there will always be a contribution from excited states, making it difficult
to reveal the true nature of the ground state in the weak-coupling regime.

\section{\label{sec:QMC} Quantum Monte Carlo method}

Since the extension of the QMC algorithm to higher dimensions is straight
forward, here we shall merely give an overview of the method which has been
discussed in detail in I.

\subsection{QMC algorithm}\label{sec:algorithm}

The cornerstone of the new approach is the canonical Lang-Firsov
transformation,\cite{LangFirsov} which separates the polaron effects, due to
the electron-phonon interaction, from the zero-point fluctuations of the
harmonic oscillators in Eq.~(\ref{eq:holstein}). The transformed model with
one electron takes the form\cite{HoEvvdL03}
\begin{equation}\label{eq:Htrans}
  \tilde{H}
  =
  \tilde{K} + P - \Ep\,,\quad
  \tilde{K}
  =
  -t \sum_{\las ij\ras} c^\dag_i c^{\phantom{\dag}}_j e^{\rmi\gamma(\op_i-\op_j)}
\end{equation}
with $P$ as defined in Eq.~(\ref{eq:holstein}), and the polaron binding
energy $\Ep=\lambda W/2$. The parameter $\gamma$, which corresponds to the
displacement of the harmonic oscillator in the presence of an
electron,\cite{HoEvvdL03} is given by $\gamma^2 = 2\Ep/\om$.  The method
employs a Trotter decomposition of the imaginary time axis into $L$ intervals
of size $\dtau = \beta/L$, where $\beta=(k_\text{B}T)^{-1}$ is the inverse
temperature. The partition function, obtained by integrating out the phonon
coordinates,\cite{HoEvvdL03} is given by
\begin{equation}\label{eq:Z}
  \Z_L
  =
  \text{const.}\int\,\D p\,\wb\,\wf\,,
\end{equation}
where $\int\D p$ denotes the $L\times N^\rD$ dimensional integral over all
phonon momenta $p$, and $N$ is the linear size of the lattice in D
dimensions. The bosonic weight $\wb$ is defined as $e^{-\dtau S_\text{b}}$
with the bosonic action
\begin{equation}\label{eq:action-w-matrix}
  S_\text{b}
  =
  \sum_{i=1}^N \bm{p}_i^\text{T} A \bm{p}_i\,,
\end{equation}
$\bm{p}_i=(p_{i,1},\dots p_{i,L})$, and a tridiagonal $L\times
L$ matrix $A$ with nonzero elements
\begin{equation}\label{eq:matrixA}
  A_{jj}
  =
  \frac{\om}{2}+\frac{1}{\om\dtau^2}\;,\quad
  A_{j,j\pm1}
  =
  -\frac{1}{\om\dtau^2}\,,
\end{equation}
and periodic boundary conditions $L+1\equiv1$.
As discussed in I, the electronic weight $\wf$ is given by
\begin{equation}\label{eq:wf}
  \wf
  =
  \text{Tr}_\text{f}\,\Omega
  \,,\quad
  \Omega
  =
  \prod_{\tau=1}^L e^{-\dtau\tilde{K}_\tau}
  \,.
\end{equation}
Here $\tilde{K}_\tau$ is $\tilde{K}$ with the phonon operators $\op_i$
replaced by the values $p_{i,\tau}$ on the $\tau$th Trotter slice.  The
exponential of the hopping term can be written as
\begin{eqnarray}\label{eq:matrices}
  e^{-\dtau\tilde{K}_\tau}
  &=&
  D_\tau \kappa D_\tau^\dag
  \\\nonumber
  \kappa_{jj'}
  &=&
  \left(e^{\dtau t\,h^\text{tb}}\right)_{jj'}\,,\quad
  (D_\tau)_{jj'}
  =
  \delta_{jj'}e^{\rmi\gamma p_{j,\tau}}\,,
\end{eqnarray}
where $h^\text{tb}$ is the $N^\rD\times N^\rD$ tight-binding hopping matrix
for the lattice under consideration. In fact, this is the only nontrivial
change compared to the one-dimensional case. As pointed out in I, for a
single electron, the fermionic trace can easily be calculated from the matrix
representation of $\Omega$ as $\wf=\sum_i\Omega_{ii}$. Due to the
complex-valued hopping term, $\wf$ is not strictly positive, which gives rise
to the minus-sign problem discussed below.  We would like to mention that, in
contrast to some determinant QMC methods,\cite{wvl1992} the $L$-fold matrix
product involved in the calculation of the matrix $\Omega$ is well
conditioned also for large systems at low temperatures, so that a
time-consuming numerical stabilization is not necessary.

In I, we have introduced the so-called principal component representation for
the phonon degrees of freedom. In terms of the latter, the bosonic weight
takes the simple Gaussian form
\begin{equation}\label{eq:action_quad}
  \wb
  =
  e^{-\dtau\sum_i \bm{\xi}^\text{T}_i\cdot\bm{\xi}_i}
\end{equation}
with the {\it principal components} $\bm{\xi}_i=A^{1/2}\bm{p}_i$. It is
strictly positive.  Finally, the efficiency of the QMC algorithm can be
greatly improved by the use of a reweighting of the probability distribution.
In our case, this amounts to transferring all the influence of the electronic
degrees of freedom---which are treated exactly---to the observables
calculated via
\begin{equation}\label{eq:reweighting}
  \las O \ras
  ~=~
  \frac{\las O\, \wf\ras_\text{b}}{\las \wf\ras_\text{b}}
\end{equation}
with the expectation values with respect to the bosonic weight $\wb$ defined
as
\begin{equation}\label{eq:Ob}
  \las O \ras_\text{b}
  =
  \frac{\int\,\D p\,\wb\; O(p)}{ \int\,\D p\,\wb}\,.
\end{equation}
As outlined in I, the expectation values defined in Eq.~(\ref{eq:Ob}) can be
determined during the QMC simulation.  The applicability of the reweighting
method has been discussed in detail in I, and here we only mention that we
have carefully checked that it also works reliably in higher dimensions. This
could be expected, since the physics of the Holstein polaron is rather
similar in all dimensions (see Sec.~\ref{sec:holstein}). Details about the
calculation of observables within our formalism can be found in I.  Due to
the analytic integration over the phonon coordinates $x$ used here,
interesting observables such as the correlation functions $\sum_i \las\on_i
\ox_{i+\delta}\ras$ are difficult to measure accurately. Other quantities
such as the quasiparticle weight, and the closely related effective
mass,\cite{WeRoFe96} can be determined from the one-electron Green function
at long imaginary times,\cite{BrCaAsMu01} but results but would not be as
accurate as existing work on the one-electron case considered here (see, \eg,
Refs.~\onlinecite{JeWh98,RoBrLi99III}, and~\onlinecite{KuTrBo02}). Therefore,
we have restricted ourselves to the kinetic energy of the electron, which
contains a lot of information about the small-polaron cross over.  For the
more demanding many-electron case, to which our method can be extended (see
Sec.~\ref{sec:summary}), other methods produce far less reliable data.

We are now in a position to discuss the actual QMC procedure which has been
explained more thoroughly in I. From the above discussion and I, it is
obvious that we only simulate the phonons, while the electronic degrees of
freedom have been integrated out analytically. This is equivalent to the
method proposed by Blankenbecler \etal\cite{BlScSu81} for the many-electron
case. However, here the Monte Carlo sampling is independent of the fermionic
weight $\wf$ given by Eq.~(\ref{eq:wf}), with the latter entering the
calculations only via the reweighting procedure [see
Eq.~(\ref{eq:reweighting})]. Thereby, we avoid the problem of nonpositive
weights during the QMC updates, since we do not use $\wf$ as a probability
for accepting or rejecting new phonon configurations. Nevertheless, the sign
problem still manifests itself in terms of statistical errors, as can be seen
from Eq.~(\ref{eq:reweighting}). Similar to other occurrences of the
minus-sign problem, \eg, for the case of the Hubbard model away from half
filling,\cite{wvl1992} simulations would become very difficult if
$\las\wf\ras_\text{b}$ should tend to zero.

Owing to the Gaussian form of the bosonic action, when written in terms of
the principal components $\xi$, the latter can be sampled exactly by drawing
random numbers from a normal distribution. In contrast to usual Markov chain
Monte Carlo simulations, every new configuration is accepted, and no
autocorrelations between successive values of the $\xi$ exist. This is in
strong contrast to conventional QMC methods for the Holstein model, also with
many electrons (see discussion in I).  Except for situations in which the
phonons are integrated out analytically, simulations become extremely
difficult at low temperatures and for small phonon frequencies due to
strongly increasing autocorrelations. We regard the complete absence of such
correlations as a major advantage of our method. After the $\xi$ have been
updated for each site of the $L\times N^\rD$ space-time lattice, a
transformation back to the momenta $p$ is performed using the matrix
$A^{-1/2}$. Then, for each observable of interest, $O\wf$ as well as the
fermionic weight $\wf$ are calculated for the current phonon configuration,
after which the next update can begin. At the end of the program, results for
observables are obtained according to Eq.~(\ref{eq:reweighting}) (see also
I).

\subsection{Sign problem}\label{sec:sign}

Before we come to a discussion of the sign problem in the approach presented
here, we would like to give a quick review of its occurrence in other QMC
methods. The situation is best illustrated for the case of the world-line
algorithm (see, \eg, Ref.~\onlinecite{batscal}). The use of the latter to
simulate systems of interacting fermions is restricted to one dimension by
the Fermi statistics of the electrons. This is a consequence of the negative
matrix elements $w$, which appear when two fermion world lines wind around
each other one or more times as they traverse the space-time lattice.  With
other methods, despite the occurrence of the sign problem, simulations can
still be carried out in many situations. Since the QMC sampling requires
nonnegative weights, one uses $|w|$ instead of $w$.  As a consequence, the
sign of the fermionic weight, $\text{sgn}(w)$, is treated as part of the
observables. The latter then have to be calculated via
\begin{equation}\label{eq:obs_sign}
  \las O\ras
  =
  \frac{\las O\,\text{sgn}(w)\ras_{|w|}}{\las \text{sgn}(w)\ras_{|w|}}\,,
\end{equation}
and it is obvious that simulations will become extremely difficult if the
denominator in Eq.~(\ref{eq:obs_sign}) tends to zero. In fact, if a sign
problem is present, the number of measurements during a QMC run has to be
increased by a factor $\propto\sign^{-2}$ to obtain results of the same
accuracy.\cite{wvl1992} As discussed in
Ref.~\onlinecite{ChWi99}, using $|w|$ instead of $w$ corresponds to
simulating an effective model of hard-core bosons. The average sign can be
written as
\begin{equation}\label{eq:sign_free_energy}
  \sign 
  =
  e^{-\beta V (f_w - f_{|w|})}\,,
\end{equation}
where $f_w$ and $f_{|w|}$ denote the free energy per site of the fermionic
and bosonic model, respectively, and $V$ is the volume of the
system.\cite{ChWi99} From Eq.~(\ref{eq:sign_free_energy}), it is obvious that
$\sign$ decreases exponentially with increasing $\beta$ and $V$.

The auxiliary-field method\cite{wvl1992} for the Hubbard model faces similar
problems. Here the weight of a configuration is given by the product of
determinants for $\UP$ and $\DO$ electrons, respectively. The product is
strictly positive only for half filling, whereas simulations for other
particle densities become very demanding at low temperatures and/or for large
systems. Since at large $U$ the determinant is similar to the weight of world
lines of Hubbard-Stratonovitch variables,\cite{Hi86} this similarity of the
dependence of $\sign$ on the parameters of the system is not surprising.
Finally, there is no sign problem in determinantal grand-canonical
simulations of the Holstein model at any filling, since the coupling of
electrons and phonons is the same for both spin
directions.\cite{LoGuScWhScSu90}

The sign problem in the current approach clearly has a fundamentally
different origin, since there is only a single electron in the system, so
that no winding of world lines around each other can take place. In fact, the
negative fermionic weights are a result of the complex phase factor in the
transformed hopping term [Eq.~(\ref{eq:Htrans})]. We will see below that, in
contrast to other methods, the sign problem in our approach diminishes, \ie,
$\sign\rightarrow 1$, with increasing system size, and a detailed
investigation of its dependence on the parameters of the model and the choice
of boundary conditions will be given in Sec.~\ref{sec:results}.

\subsection{Numerical details and performance}\label{sec:numerics}

As discussed in I, the error due to the Trotter decomposition is proportional
to $(\dtau)^2$. We perform simulations at different values of $\dtau$,
typically $\dtau = 0.1$, $0.075$ and $0.05$, and exploit the linear
dependence of the results on $(\dtau)^2$ to extrapolate to $\dtau=0$. This is
a common procedure in the context of discrete-time QMC methods,\cite{wvl1992}
and allows one to remove the Trotter error if the values of $\dtau$ are
sufficiently small. Moreover, a given accuracy can often be reached using
larger values of $\dtau$, compared to calculations which do not use the
abovementioned extrapolation, thereby saving computer time.

We conclude this section by a comparison of our approach with other QMC
methods for the Holstein polaron. As mentioned in
Sec.~\ref{sec:introduction}, the methods of de Raedt and
Lagendijk\cite{dRLa82,dRLa83,dRLa84} and Kornilovitch\cite{Ko97,Ko98,Ko99}
are based on an analytic integration over the phonon degrees of freedom. This
separation of electronic and bosonic degrees of freedom greatly reduces the
statistical noise due to phonon fluctuations, which are induced by the
electron-phonon interaction. The fluctuations increase noticeably with
decreasing phonon frequency, decreasing temperature and increasing
electron-phonon coupling. Although the Lang-Firsov transformation used here
performs a very similar task, namely, to separate polaron effects from the
zero-point and thermal fluctuations of the free oscillators (see I), the
integral over the bosonic degrees of freedom is calculated with Monte Carlo,
thereby leaving us with a residual influence of the phonons. In fact, the
numerical effort for calculations with our approach is proportional to the $L
N^{3\rD}$, similar to the grand-canonical determinant method for the Holstein
model.\cite{BlScSu81} This could be improved to a computer time $\sim
N^{2\rD}$ by linearizing the exponential of the hopping term to a tridiagonal
matrix, which has not been done in the work presented here.  Although our
algorithm is not as efficient as the methods of
Refs.~\onlinecite{dRLa82,dRLa83,dRLa84,Ko97,Ko98,Ko99}, in which the
numerical effort is independent of $N$, proportional to $L^2$ and depends
only linearly on D, we will see in the following section that it is well
suited for accurate simulations on large clusters in one and two dimensions,
and on reasonably large clusters in 3D.  Moreover, as we are interested in
developing a QMC method that can also be applied to the many-electron system
in the adiabatic regime---in order to study, \eg, quantum phonon effects in
the manganites (see I)---such a decrease in performance when compared to the
world-line methods\cite{dRLa82,dRLa83,dRLa84,Ko97,Ko98,Ko99} is acceptable,
since the latter cannot be applied to the many-electron case in more than one
dimension. Additionally, the advantage of the analytical integration over the
phonons is biggest in the one-electron case, in which they represent the
majority of the degrees of freedom in the path-integral representation of the
partition function.\cite{Ko97} This is no longer true at finite band filling,
where the contributions of fermions and bosons are comparable.

\section{\label{sec:results} Results}

This section is divided into two parts. First, in
Sec.~\ref{sec:sign-problem}, we investigate in detail the dependence of the
aforementioned sign problem on electron-phonon coupling, dimensionality,
boundary conditions, system size, phonon frequency, and temperature. In
Sec.~\ref{sec:polaron}, we present results for the electronic kinetic energy
to study the small-polaron cross over discussed in Sec.~\ref{sec:holstein}.

\subsection{Sign problem}\label{sec:sign-problem}

\begin{figure}
  \includegraphics[width=0.45\textwidth]{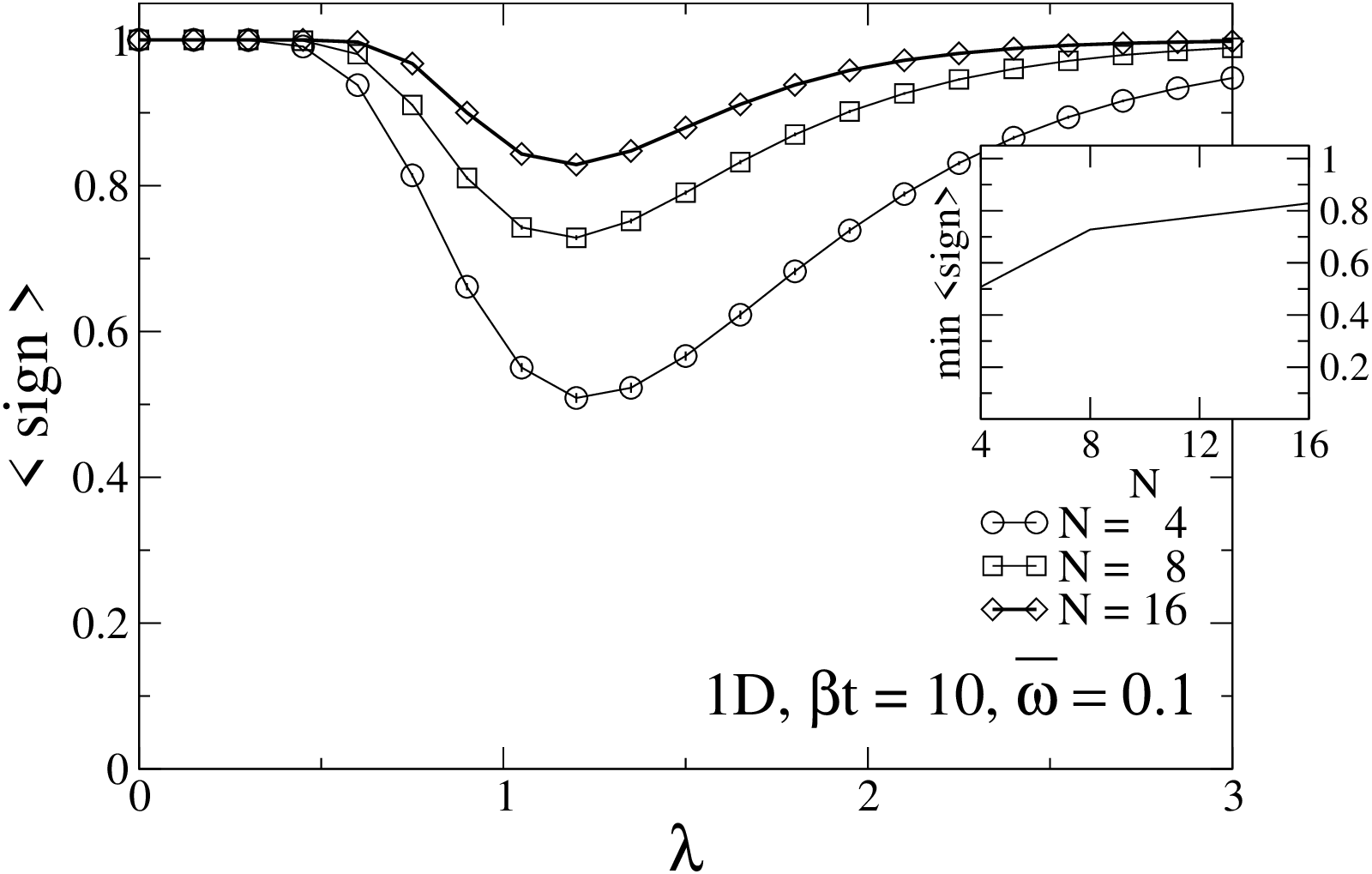}
\caption{\label{fig:sign_1d_N}
  Average sign of the fermionic weight $\wf$ as a function of electron-phonon
  coupling $\lambda$ in one dimension for different cluster sizes $N$, with
  parameters as indicated in the figure. Here and in subsequent figures,
  lines are guides to the eye only, and error bars are smaller than the
  symbols shown. The data presented in
  Figs.~\ref{fig:sign_1d_N}\,--\,\ref{fig:sign_3d_N} are for $\dtau=0.05$.
  The inset shows the minimum of $\sign$ as a function of the system size $N$
  (see text).
}
\end{figure}
\begin{figure}
  \includegraphics[width=0.45\textwidth]{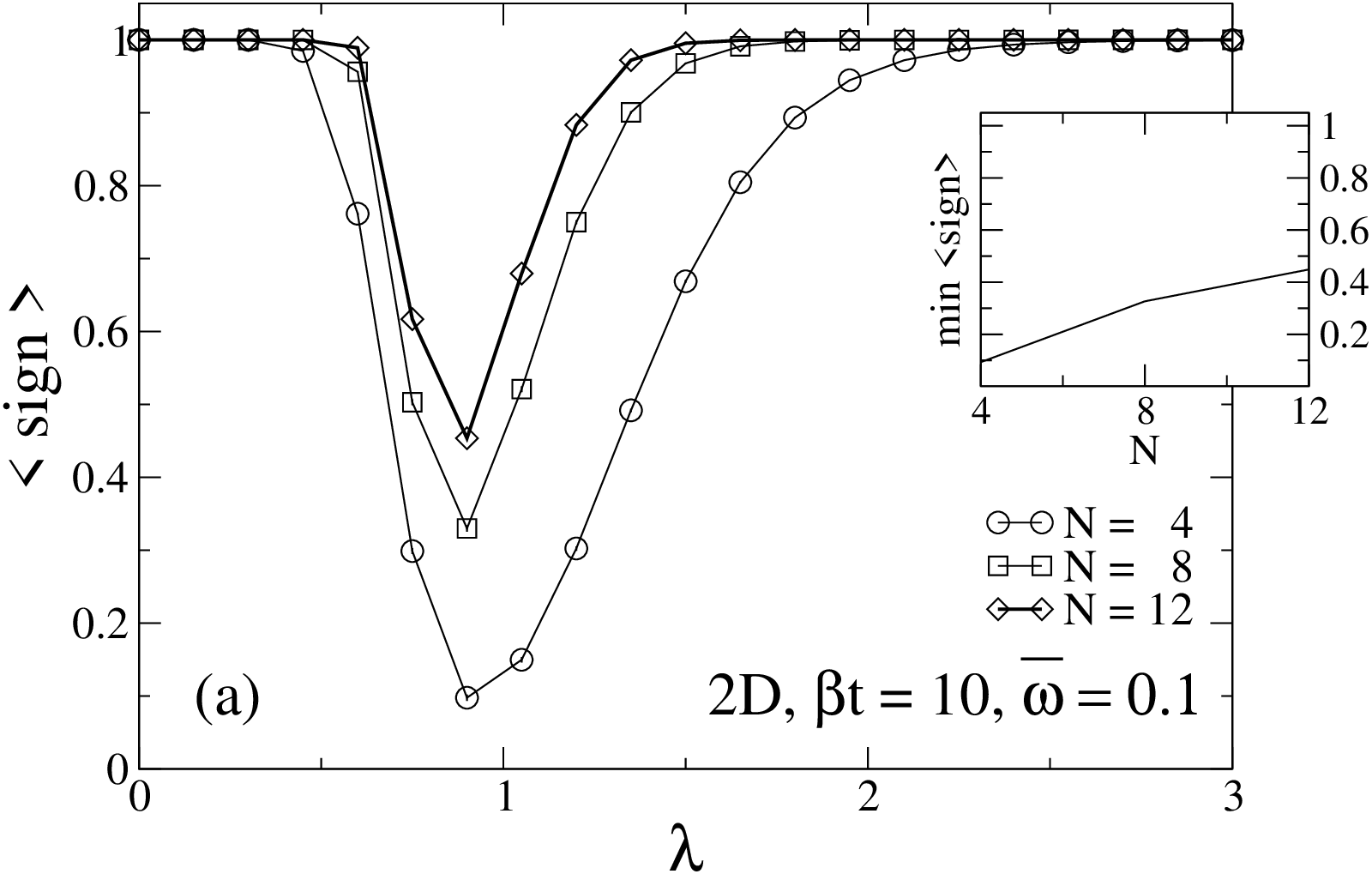}\\
  \includegraphics[width=0.45\textwidth]{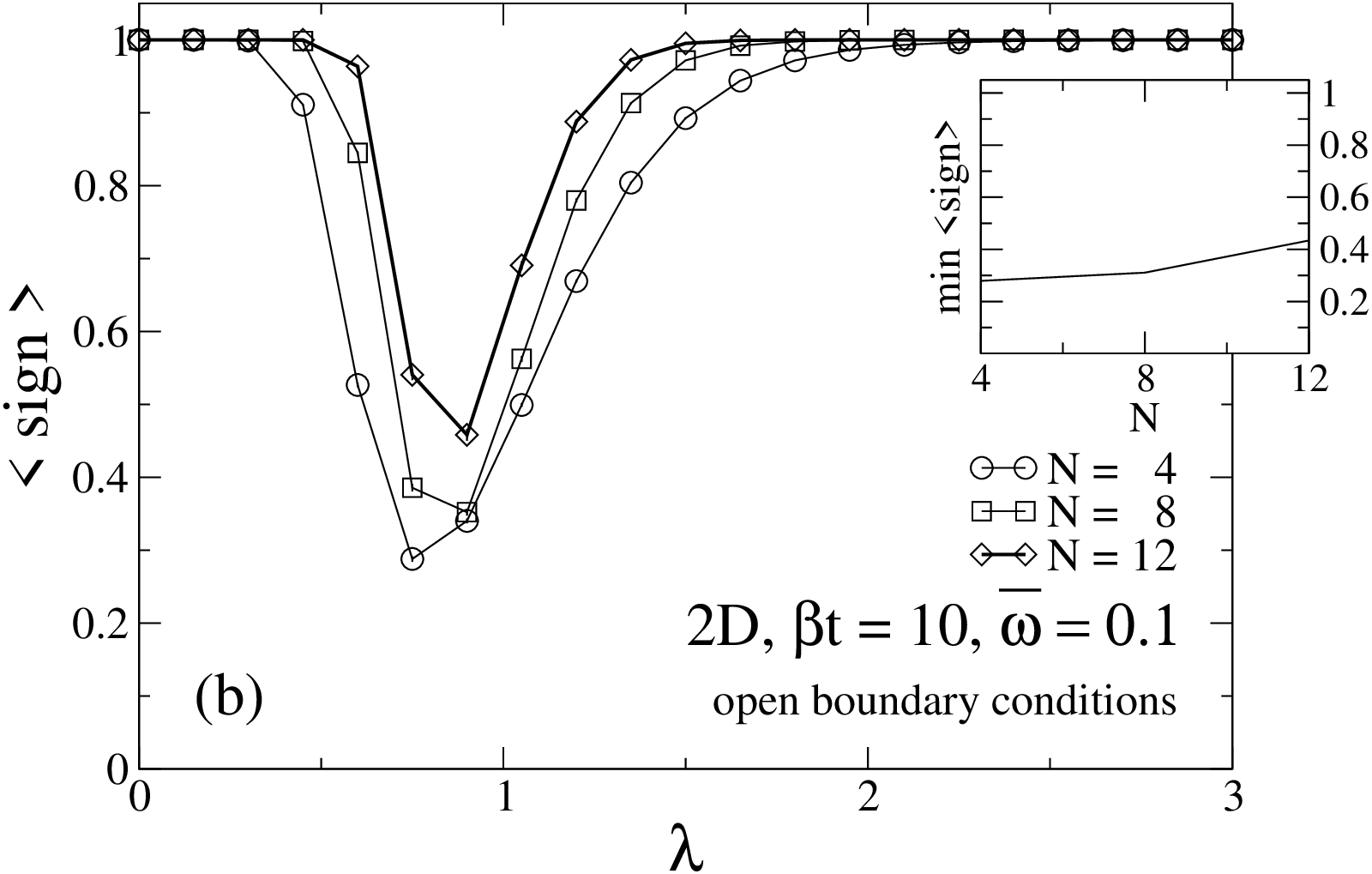}
\caption{\label{fig:sign_2d_N}
  As in Fig.~\ref{fig:sign_1d_N}, but for the case of two-dimensional
  clusters with (a) periodic boundary conditions and (b) open boundary
  conditions in real space. The insets show the minimum of $\sign$ as a
  function of the linear system size $N$.}
\end{figure}
\begin{figure}
  \includegraphics[width=0.45\textwidth]{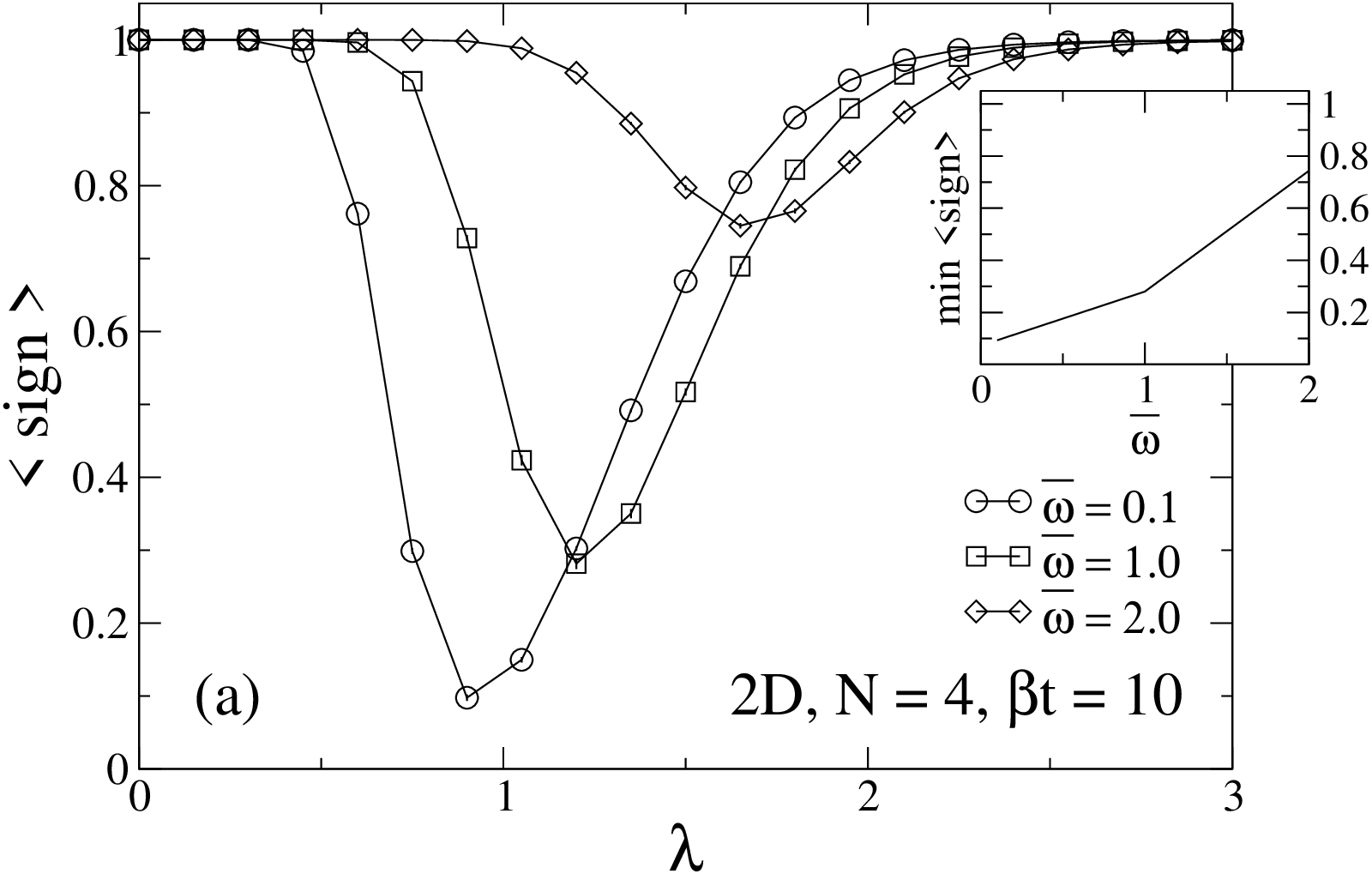}\\
  \includegraphics[width=0.45\textwidth]{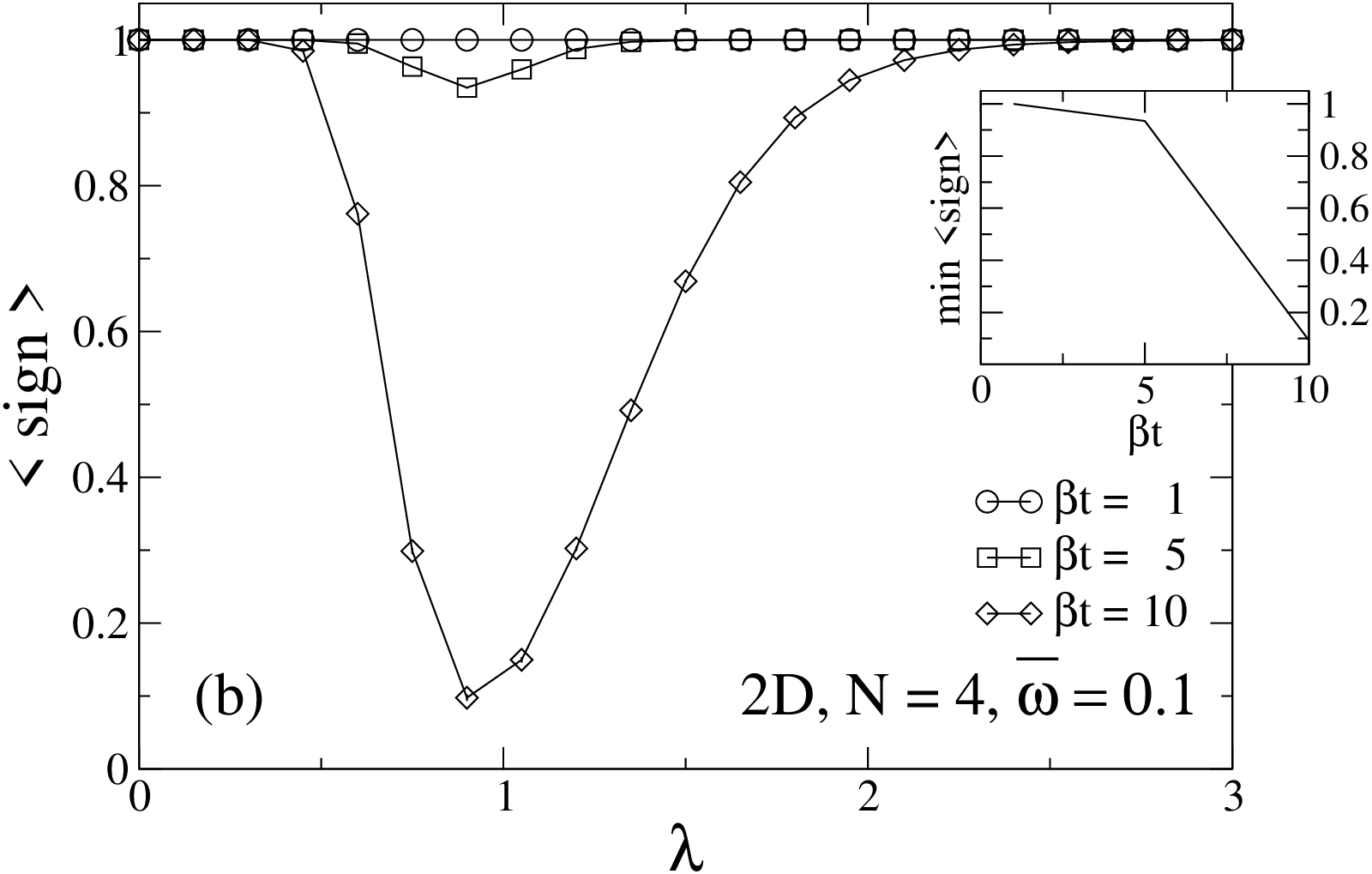}
\caption{\label{fig:sign_2d_par}
  Dependence of the average sign of $\wf$ on (a) phonon frequency
  $\omb$ and (b) inverse temperature $\beta$ on a $4\times4$ cluster.
  The insets show the minimum of $\sign$ as a function of (a) phonon
  frequency $\omb$, and (b) inverse temperature $\beta t$. }
\end{figure}
\begin{figure}
  \includegraphics[width=0.45\textwidth]{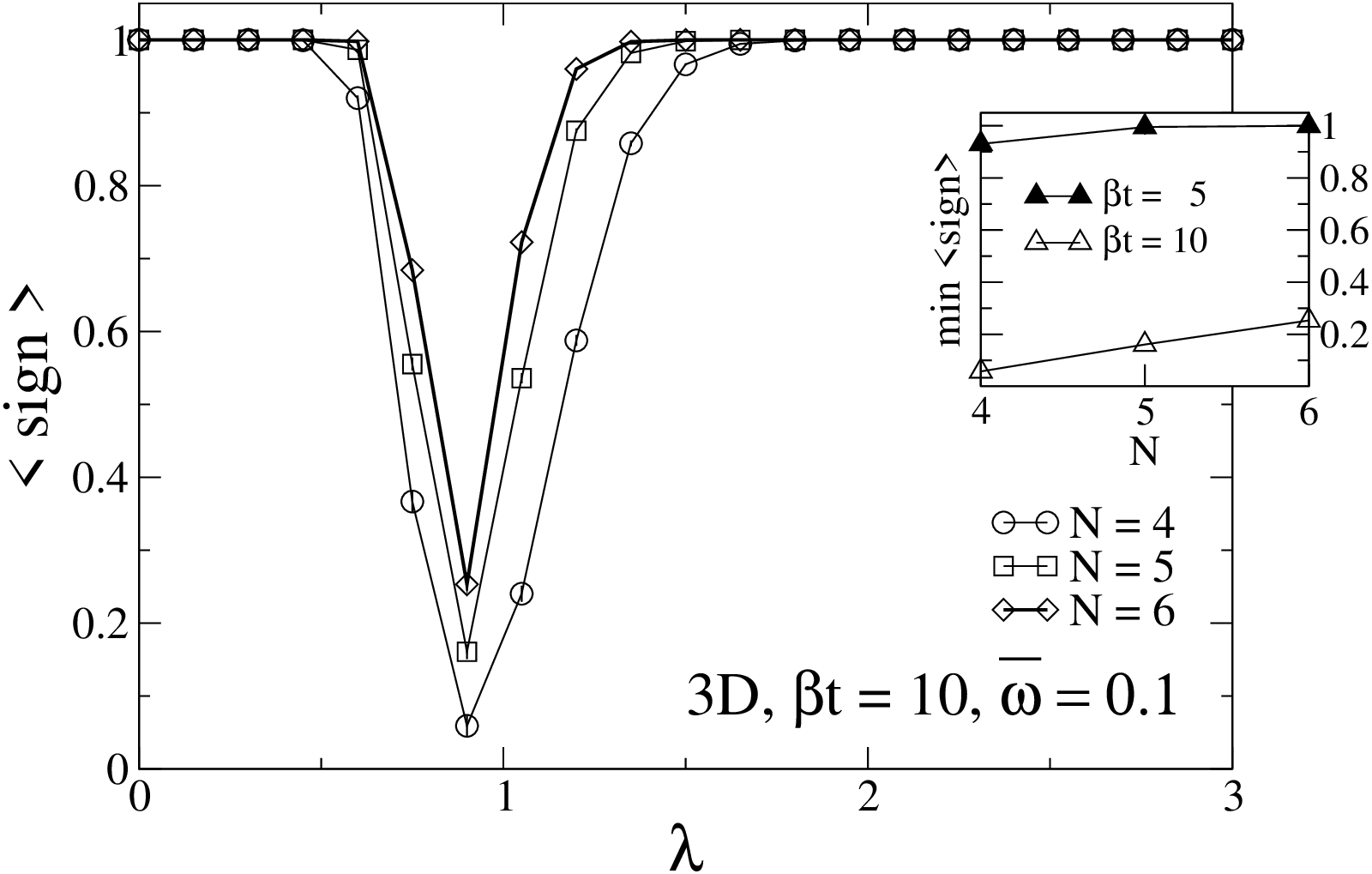}\\
\caption{\label{fig:sign_3d_N}
  Dependence of the average sign of $\wf$ on the linear system size $N$ in
  three dimensions. The inset shows the minimum of $\sign$ as a function of
  the linear system size $N$ for two different temperatures.}
\end{figure}

Here we are interested in the average sign of the fermionic weight $\wf$,
which is given by
\begin{equation}
  \sign
  =
  \frac{\las\wf\ras_\text{b}}{\las|\wf|\ras_\text{b}}\,,
\end{equation}
where the expectation value $\las\cdots\ras_\text{b}$ has been defined in
Eq.~(\ref{eq:Ob}). In Fig.~\ref{fig:sign_1d_N}, we show its dependence on
electron-phonon coupling and system size. We would like to point out that
while the extrapolation to $\dtau=0$, discussed in Sec.~\ref{sec:QMC}, has
been used for the results shown in the following subsection, the calculations
for the average sign have been performed for a single, fixed value
$\dtau=0.05$, for which the Trotter error is smaller than statistical errors.
Some other results for $\sign$ in one dimension have already been reported in
I.

From the general discussion in Sec.~\ref{sec:QMC}, it is clear that the sign
problem encountered in the present approach is of a different nature than in,
for example, QMC simulations of the Hubbard model. As reported in I, for the
Holstein polaron problem under consideration, it is most pronounced for {\it
  small} systems, low temperatures and small phonon frequencies $\omb\ll1$.
Therefore, the bulk of results presented below will be for such a set of
``worst case'' parameters, including $N=4$, $\beta t=10$, and $\omb=0.1$.
Figure~\ref{fig:sign_1d_N} shows that, in one dimension, the average sign of
$\wf$ in the critical region of intermediate electron-phonon coupling
increases quickly as the system size increases from $N=4$ to $N=16$, which is
in strong contrast to Eq.~(\ref{eq:sign_free_energy}). This increase of the
minimum as a function of $N$ is also shown in the inset of
Fig.~\ref{fig:sign_1d_N}. As we have only calculated $\sign$ for a finite
number of $\lambda$-values, an approximation for the minimum has been
determined using a spline interpolation.

The minimum of $\sign$ occurs near $\lambda=1$, where the transition from a
large to a small polaron takes place (see Sec.~\ref{sec:holstein}). The
increase of statistical errors in this regime as a consequence of the sign
problem is similar to the situation encountered with simulations of the
untransformed model (see I). However, the use of the transformed model still
gives significantly more accurate results for the same number of
measurements, in particular for low temperatures and small phonon
frequencies.  Finally, one may be tempted to explain the unusual system-size
dependence of the sign problem by ascribing its origin to the periodic
boundary conditions in real space. If the latter were indeed the source of
the sign problem, the boundary effects would decrease with increasing system
size, in accordance with the results of Fig.~\ref{fig:sign_1d_N}. This
possibility has been investigated, and we shall see below that the sign
problem persists also for the case of open boundary conditions.

We now come to the Holstein model in two dimensions. In
Fig.~\ref{fig:sign_2d_N}(a), we show results for $\sign$ for different
lattice sizes, again starting with a very small linear dimension $N$. All
other parameters are the same as before, in particular $\beta t=10$ and
$\omb=0.1$. Obviously, for the smallest cluster size shown, the minimum of
the average sign has diminished to a value of approximately 0.1, so that
large numbers of measurements are necessary. However, similar to one
dimension, $\sign$ increases with increasing system size, and for the largest
system size shown ($N=12$), we find a rather uncritical minimum value of
about 0.5.

The results for open boundary conditions, shown in
Fig.~\ref{fig:sign_2d_N}(b), reveal that for small clusters, the average sign
increases compared to the case of periodic boundary conditions. However, with
increasing system size, $\sign$ quickly converges to the same values,
independent of the boundary conditions. This is just what one would expect,
since with increasing $N$, the effect of the choice of boundary conditions on
the properties of the system diminishes.  Moreover, we can conclude that the
negative weights do not simply result from hopping processes of the electron
across the periodic boundaries, since in that case we would expect
$\sign\equiv1$ for open boundary conditions, in contrast to
Fig.~\ref{fig:sign_2d_N}(b). Similar behavior has been found in one and three
dimensions.

The influence of the phonon frequency $\om$ on the average sign is shown in
Fig.~\ref{fig:sign_2d_par}(a). Clearly, the sign problem is most noticeable
for small values of $\omb$, while it diminishes quickly as we increase the
adiabatic ratio (see inset). This is very similar to the small-polaron cross
over. As discussed in Sec.~\ref{sec:holstein}, the latter sharpens
significantly with decreasing $\omb$, while an abrupt transition is
completely absent in the nonadiabatic regime $\omb>1$. Moreover, the coupling
$\lc$ at which the minimum of $\sign$ occurs increases with $\omb$, in
agreement with the aforementioned small-polaron condition
$\lambda\rD/\omb>1$.

In Fig.~\ref{fig:sign_2d_par}(b), we report the average sign as a function of
$\lambda$, and for different inverse temperatures $\beta$. Again we have
taken $N=4$ and $\omb=0.1$, the parameters for which the sign problem is most
noticeable. While for $\beta t=10$, the minimum of $\sign$ lies below 0.1,
the situation is much better already for $\beta t=5$, as shown in the inset.
At even higher temperature $\beta t = 1$, the fermionic weight is always
positive so that we have $\sign=1$ for all $\lambda$. The dependence of the
sign problem on temperature is therefore similar to other QMC methods [see
Eq.~(\ref{eq:sign_free_energy})], although we do not find a simple
exponential relation.

Finally, we also present results for $\sign$ in three dimensions, for
lattices of different linear size $N$, and again for the parameters
$\omb=0.1$ and $\beta t=10$ as a function of $\lambda$.
Figure~\ref{fig:sign_3d_N} reveals that the minimum of $\sign$ in 3D has an
even more pronounced form than in two dimensions. The sign problem diminishes
slightly as we increase the system size from $N=4$ to $N=6$. However,
accurate simulations in this regime are still quite demanding. As we will see
in Sec.~\ref{sec:polaron}, a temperature of $\beta t=5$ is sufficient to
study the low-temperature properties of the Holstein model. For this higher
temperature, and all other parameters unchanged, $\sign$ is close to 1 even
for $N=4$ (see inset in Fig.~\ref{fig:sign_3d_N}), so that accurate results
can be obtained even for small phonon frequencies in three dimensions (see
Sec.~\ref{sec:polaron}).

In summary, the investigation of the sign problem has shown that our method
works well for a large range of values of phonon frequency, electron-phonon
coupling and temperature, as long as the system size is large enough. This
contrasts, for example, with the world-line algorithm of
Kornilovitch,\cite{Ko98,Ko99} which is restricted to intermediate and strong
coupling, as well as low temperatures and rather large phonon frequencies
(see also Sec.~\ref{sec:holstein}), also by a minus-sign problem. Although we
have investigated the influence of all important parameters on the sign
problem, a physical interpretation of its origin has not emerged.
Nevertheless, it is clear that the negative fermionic weights are a result of
the phase factors in the Lang-Firsov transformed hopping term in
Eq.~(\ref{eq:Htrans}). This is similar to the sign problem which occurs, \eg,
in simulations of electron-phonon models in an external magnetic
field,\cite{LoGuScWhScSu90} but here the phonon fields $p_{i,l}$ vary with
time ($l$) and position ($i$), and are coupled in imaginary time by the
bosonic action [see Eqs.~(\ref{eq:action-w-matrix}) and~(\ref{eq:matrixA})].
Moreover, the dependence of the sign problem on $\omb$, $\lambda$, $\beta$
and $N$ bears striking resemblance to the influence of these parameters on
the properties of the Holstein polaron. In particular, its reduction with
increasing system size may be a consequence of the dilution of the system
(the particle density $n\rightarrow 0$ as $N\rightarrow\infty$, in the
one-electron case). Finally, it is interesting to note that the large
statistical fluctuations---resulting from the sign problem in the case of the
transformed model---occur at exactly the same points in parameter space as in
the untransformed model. This suggests a strong correlation between the
minimum in $\sign$ and the transition to a small polaron, which both occur at
or near $\lc$, similar to simulations of other models, in which the sign
problem occurs exactly where the most interesting physics is going on, \ie,
in the vicinity of phase transitions.\cite{wvl1992}

\subsection{Holstein polaron}\label{sec:polaron}

\begin{figure}
  \includegraphics[width=0.45\textwidth]{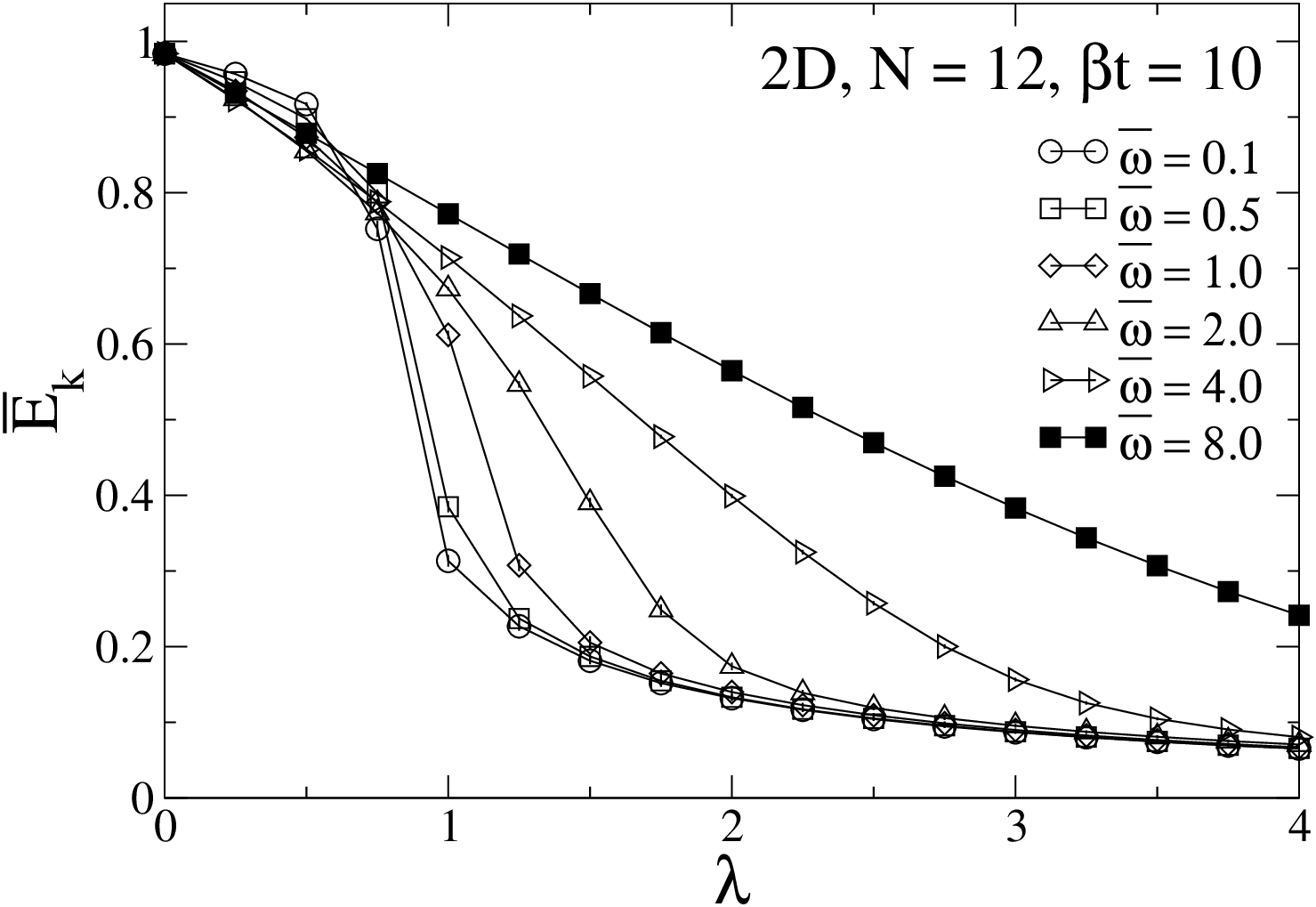}
  \caption{\label{fig:Ek_2d_various_w}
  Normalized kinetic energy $\Ek$ [see Eq.~(\ref{eq:Ek})] as a function of
  electron-phonon coupling $\lambda$ for different values of the phonon
  frequency $\om$ on a $12\times12$ lattice. The results shown in
  Figs.~\ref{fig:Ek_2d_various_w}\,--\,\ref{fig:Ek_dimensionality} have been
  obtained by extrapolating the QMC data to $\dtau=0$ (see text).}
\end{figure}

The most interesting observable, which is easily accessible with our method
(see discussion in Sec.~\ref{sec:algorithm}),
and allows us to investigate the small-polaron cross over, is the
one-electron kinetic energy $E_\text{k}=\las K\ras$, given by the expectation
value of the first term in Hamiltonian~(\ref{eq:holstein}). In order to
compare results for different dimensions, we define the normalized quantity
\begin{equation}\label{eq:Ek}
  \Ek
  =
  E_\text{k}/(-2t\rD)
\end{equation}
with $\Ek=1$ for $T=0$ and $\lambda=0$.  Due to the large amount of work that
has been devoted to the Holstein polaron in the past (see
Sec.~\ref{sec:holstein}), a lot is known about the transition from a large to
a small polaron with increasing electron-phonon coupling. In this work, we
therefore concentrate on those aspects which have not been studied in a
systematic way so far. To this end, we exploit the advantages of our QMC
method which allows us to investigate, in particular, finite-size and
finite-temperature effects. The latter have only been touched upon briefly by
previous authors employing QMC,\cite{dRLa82,dRLa83,dRLa84,Ko97,Ko98,Ko99} who
focused on very large\cite{dRLa82,dRLa83,dRLa84,Ko97} or infinite
systems,\cite{Ko98,Ko99} in the ground state\cite{Ko98,Ko99} or at two
different temperatures.\cite{dRLa82,dRLa83} However, since a large amount of
work has been done using ED or other methods based on diagonalization of
small clusters, it is essential to study the convergence of the results with
increasing system size. Additionally, we shall also present a comparison of
the kinetic energy in one, two and three dimensions.  As pointed out before,
the results shown here have been obtained by extrapolating to $\dtau=0$,
thereby eliminating the error due to the Trotter discretization.

\subsubsection{Two dimensions}

To study the small-polaron cross over, previous authors focused on the
effective mass of the electron\cite{Ko98,JeWh98,RoBrLi99III,KuTrBo02} and the
quasiparticle weight.\cite{WeFe97,CadFIa99,KuTrBo02} Unfortunately, as
pointed out in Sec.~\ref{sec:algorithm}, these observables cannot be
calculated directly with our QMC algorithm. Nevertheless, we can examine many
interesting aspects of the cross over by calculating the kinetic energy of
the electron, and comparison will be made to existing
work.\cite{dRLa82,dRLa83,dRLa84,Ko97,JeWh98}

We begin with the dependence of the cross over on the phonon frequency. To
this end, we present in Fig.~\ref{fig:Ek_2d_various_w} results for the
kinetic energy calculated for $N=12$, $\beta t=10$ and different values of
$\omb$. The large range of the adiabatic ratio in
Fig.~\ref{fig:Ek_2d_various_w}, $0.1\leq\omb\leq8.0$, shows the ability of
our method to give accurate results for almost arbitrary values of the phonon
frequency, especially in the adiabatic regime $\omb\ll1$. This is in contrast
to the method of Kornilovitch,\cite{Ko98,Ko99} which is restricted to
$\omb\gtrsim1$ and $\lambda\gtrsim1$ by a severe minus-sign problem. In our
case, the only limitations regarding the accessible values of $\omb$ are the
moderate sign problem, discussed in Sec.~\ref{sec:sign-problem}, which for
small systems and low temperatures gives rise to a noticeable increase of
statistical errors as $\omb\rightarrow0$, and the increasing Trotter error as
$\omb\rightarrow\infty$, which requires the use of more and more time slices
(see discussion in I).
\begin{figure}
  \includegraphics[width=0.45\textwidth]{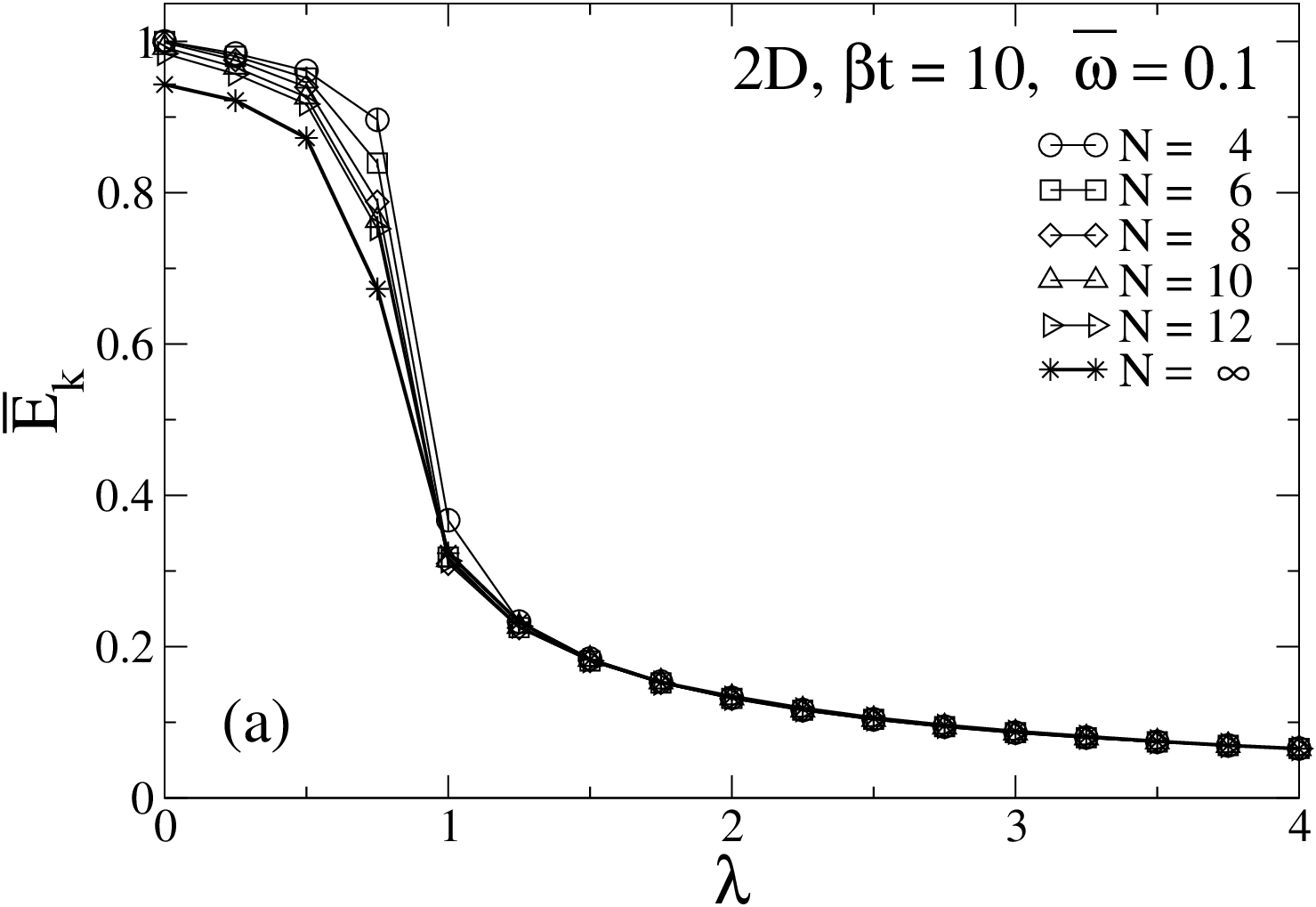}\\
  \includegraphics[width=0.45\textwidth]{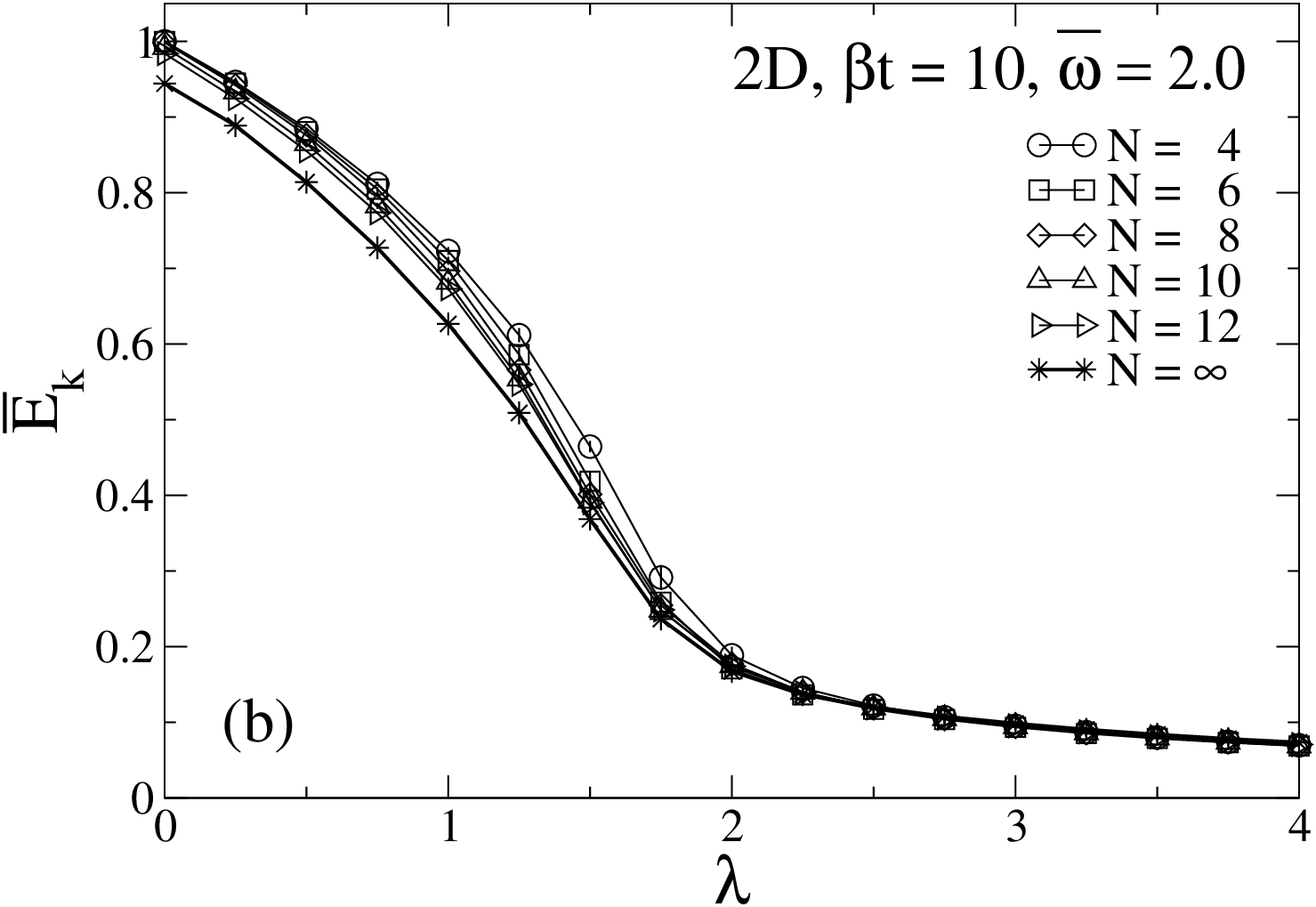}
\caption{\label{fig:Ek_2d_size}
  Normalized kinetic energy $\Ek$ as a function of electron-phonon coupling
  $\lambda$ for different linear dimensions $N$ of the system.}
\end{figure}
\begin{figure}
  \includegraphics[width=0.45\textwidth]{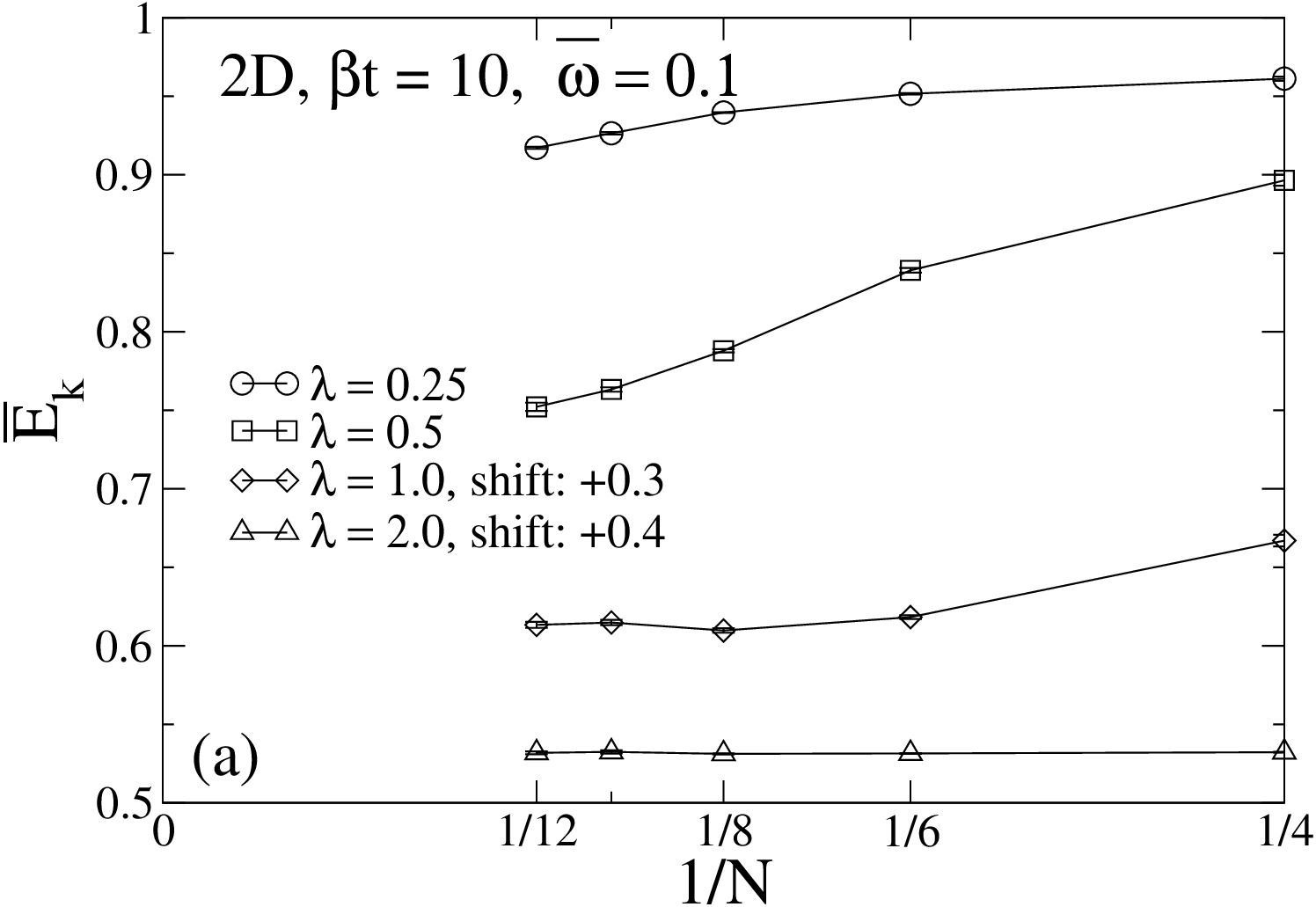}\\
  \includegraphics[width=0.45\textwidth]{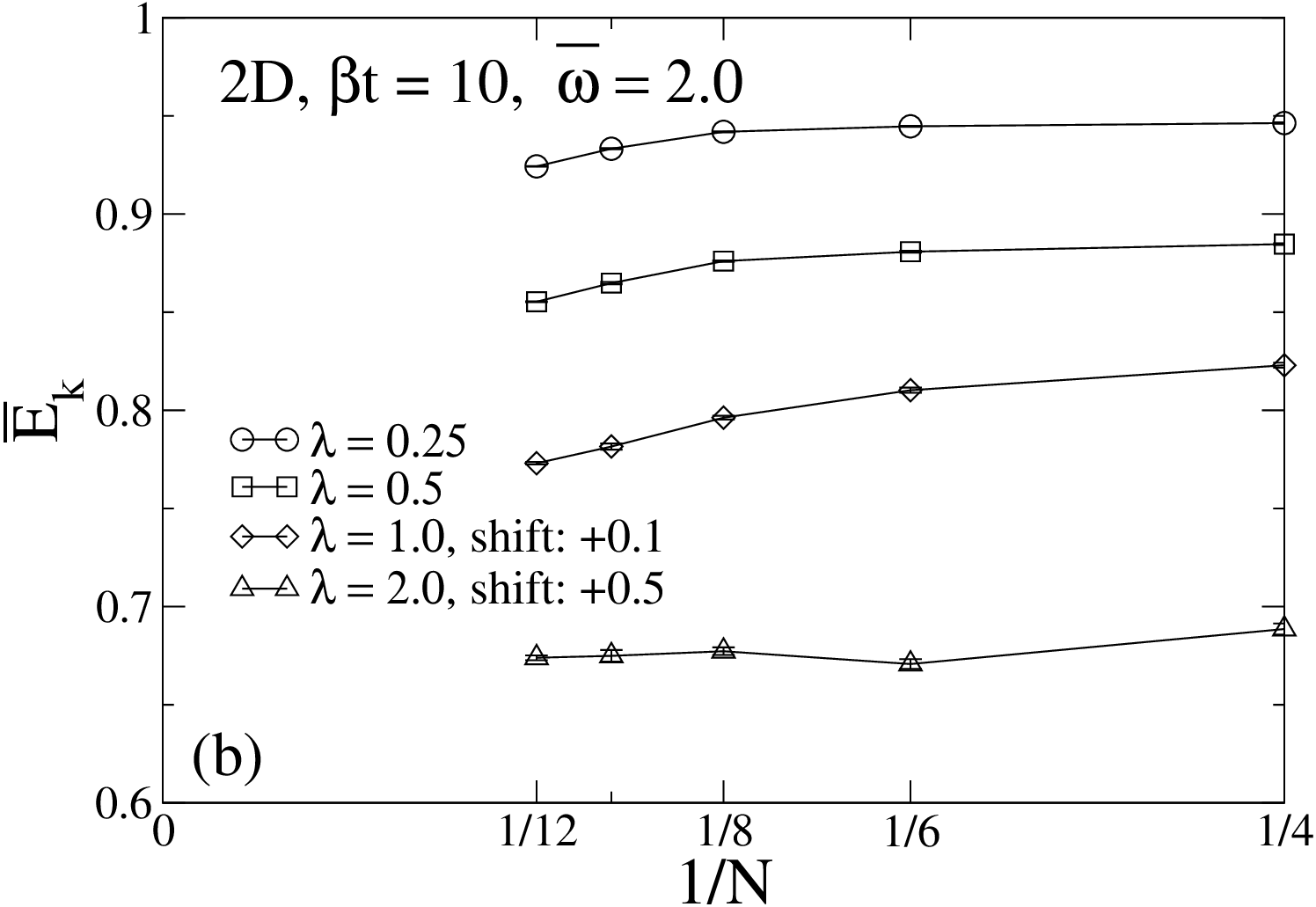}
\caption{\label{fig:Ek_finite_size}
  Normalized kinetic energy $\Ek$ as a function of the inverse of the linear
  size $N$ of the system, and for different values of the electron-phonon
  coupling $\lambda$. As indicated in the legend, some curves have been
  shifted, in order to allow for a better representation. All curves are
  monotonic within statistical errors.}
\end{figure}

\begin{figure}
  \includegraphics[width=0.45\textwidth]{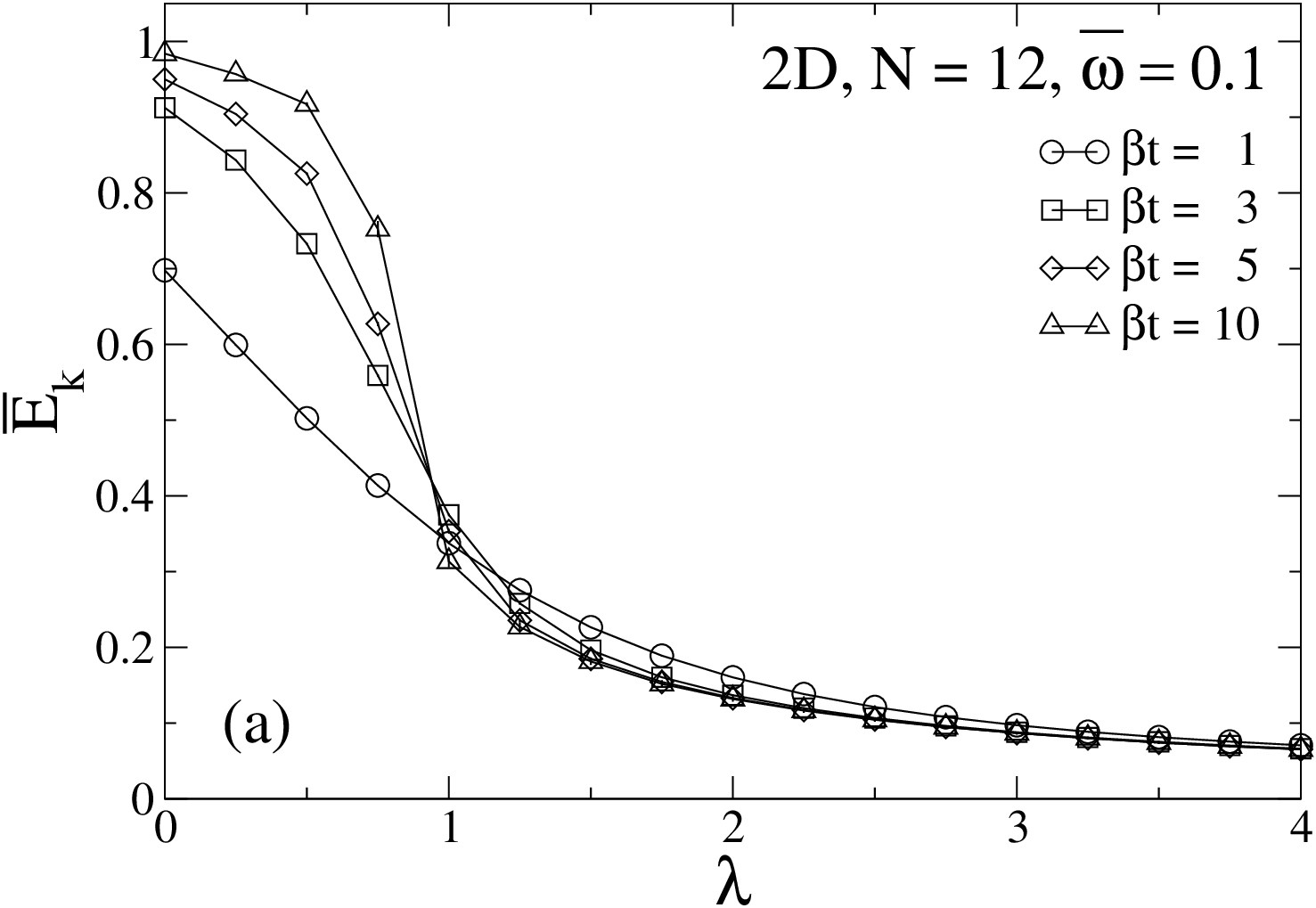}\\
  \includegraphics[width=0.45\textwidth]{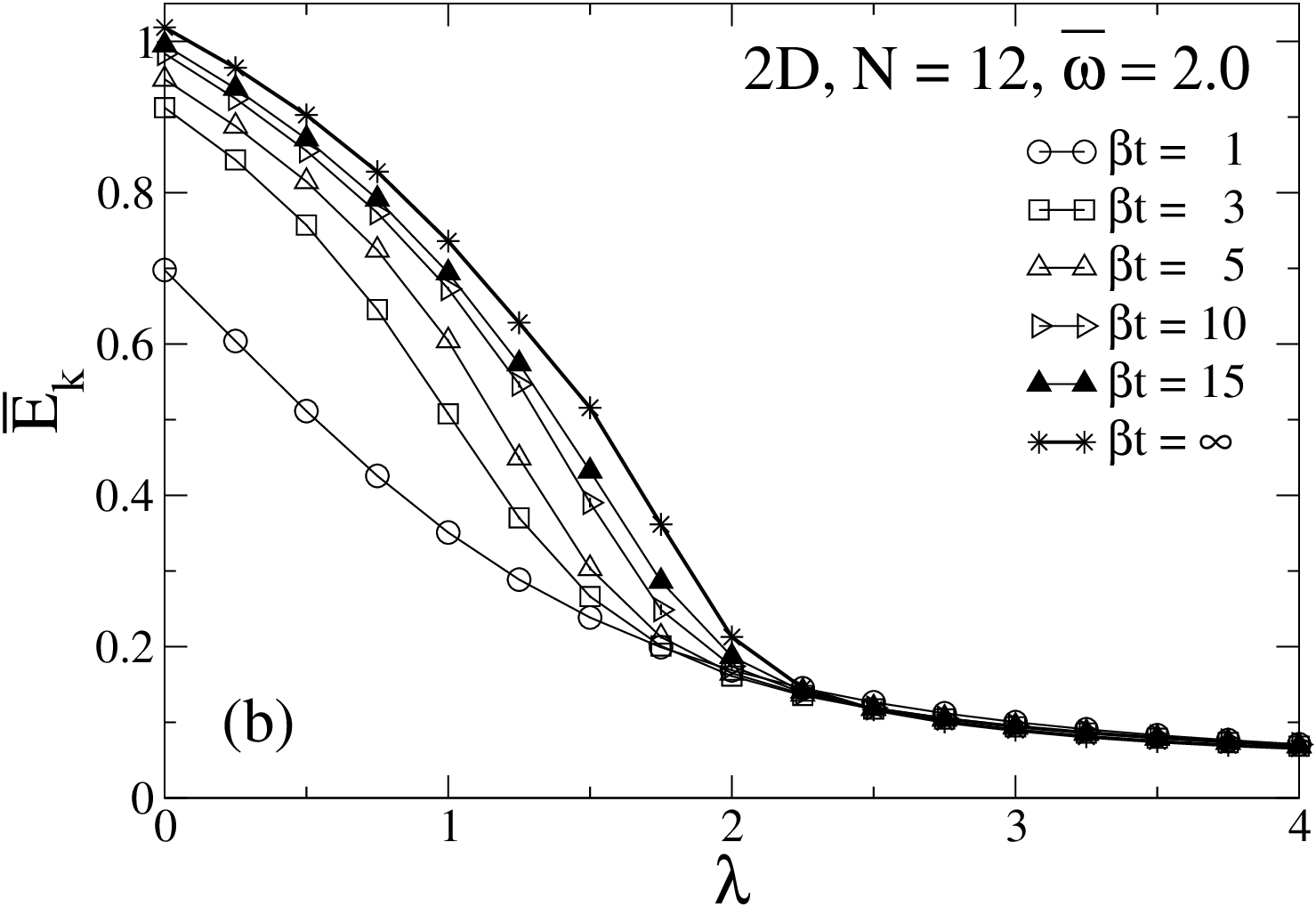}
\caption{\label{fig:Ek_2d_beta}
  Normalized kinetic energy $\Ek$ as a function of electron-phonon coupling
  $\lambda$ for different inverse temperatures $\beta$.}
\end{figure}
\begin{figure}
  \includegraphics[width=0.45\textwidth]{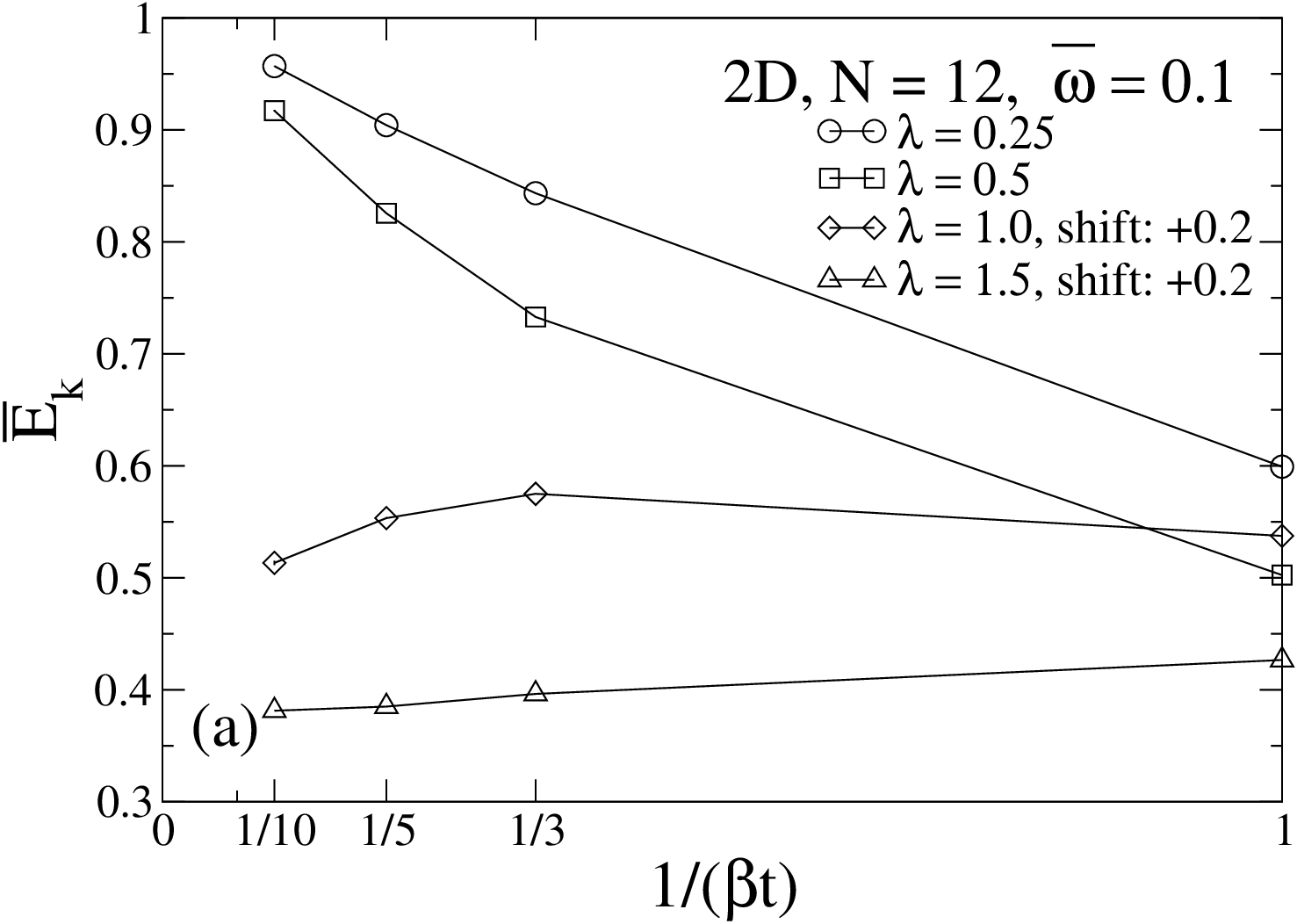}\\
  \includegraphics[width=0.45\textwidth]{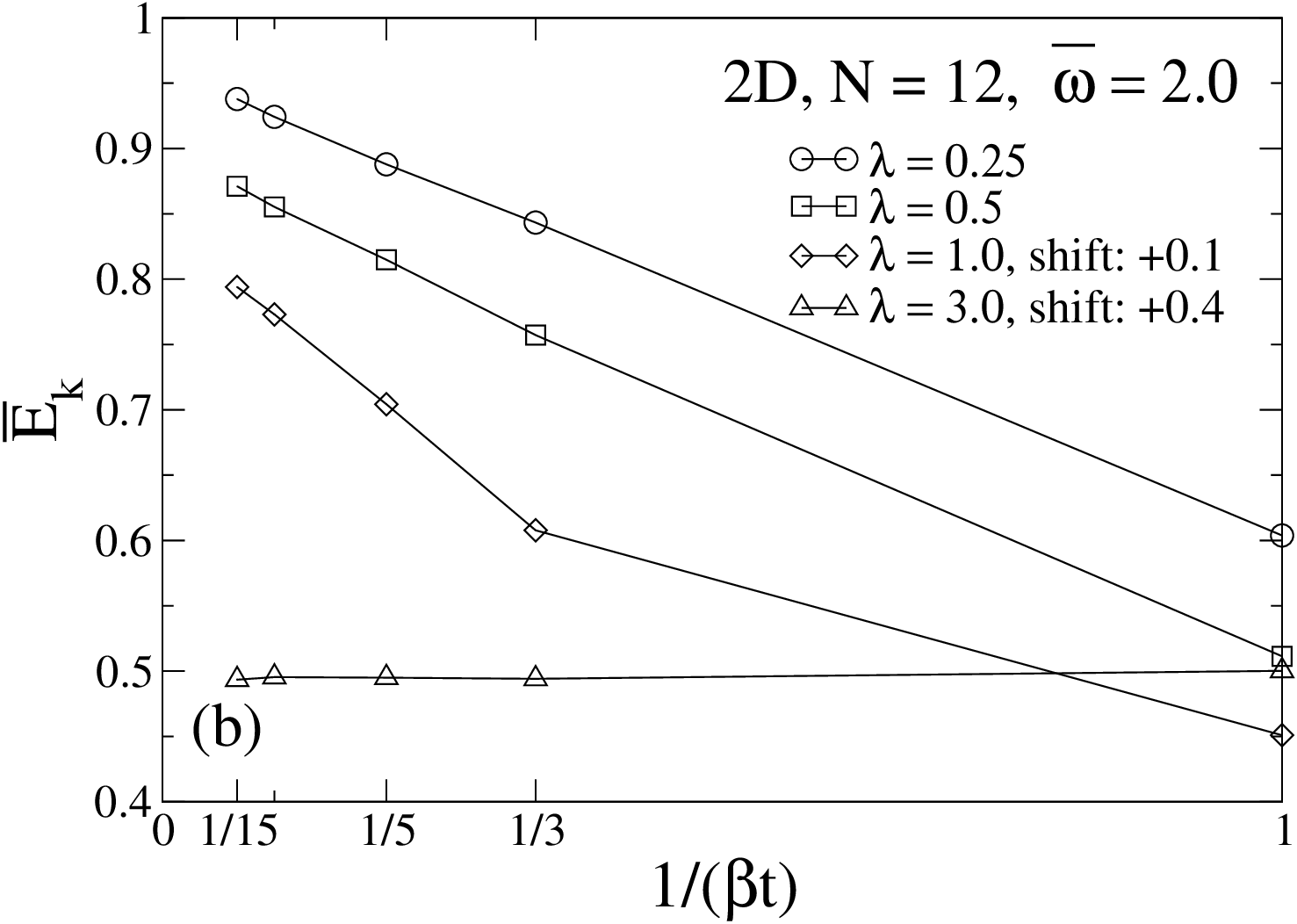}
\caption{\label{fig:Ek_2d_beta_scaling}
  Normalized kinetic energy $\Ek$ as a function of temperature $1/(\beta
  t)$, and for different values of the electron-phonon coupling
  $\lambda$. As indicated in the legend, some curves have been
  shifted, in order to allow for a better representation.}
\end{figure}
Figure~\ref{fig:Ek_2d_various_w} also shows the well known fact that the
transition from large to small polaron near $\lambda=1$ sharpens considerably
with decreasing phonon frequency. While there is an abrupt decrease in $\Ek$
for $\omb=0.1$, the cross over is very smooth for $\omb\gtrsim1$.
Additionally, we see from Fig.~\ref{fig:Ek_2d_various_w} that the cross over
position $\lc$ increases with $\omb$ in the nonadiabatic regime. The physics
of the transition to a small polaron has been discussed, \eg, by Capone
\etal\cite{CaStGr97} In the adiabatic regime ($\omb\ll1$), the cross over is
entirely determined by the balance of kinetic and electron-phonon coupling-
or displacement energy, where the latter is given by the polaron binding
energy $\Ep$. As soon as the gain in displacement energy outweighs the loss
in kinetic energy, the electron localizes in a potential well and forms a
polaron. The parameter $\lambda$ is defined as the ratio of these two
contributions and may be written as $\lambda=\Ep/(W/2)$ ($-W/2$ is the
kinetic energy of a free electron at $T=0$). Therefore, in the adiabatic
regime, the cross over occurs at $\lc=1$. With increasing $\omb$, the lattice
energy becomes more and more important, since more energy is required to
excite phonons. As a consequence, the distortions of the lattice around the
position of the electron---giving rise to the large effective mass and low
mobility in the small polaron regime---are much smaller, and the local
oscillators will predominantly be in their ground state. Thus, even for
$\lambda>1$, where a trapped state is energetically favored in the adiabatic
regime, the electron remains mobile. The decrease of $\Ek$ with increasing
$\lambda$ is a result of the exponentially decreasing overlap of a displaced
and an undisplaced harmonic oscillator in its ground state, which reduces the
hopping matrix element between neighboring lattice sites. In the nonadiabatic
regime, a small polaron is formed if $\gamma^2=\Ep/\om>1$ (see
Ref.~\onlinecite{CaStGr97}), where $\gamma$ is the induced lattice distortion
at the site of the electron which follows from the Lang-Firsov
transformation.\cite{HoEvvdL03} This is identical to the condition
$\lambda\rD/\omb>1$ given above. The larger lattice energy also gives rise to
the more gradual decrease of $\Ek$ for intermediate and large values of
$\omb$. In particular, the kinetic energy is much larger for $\lambda>1$ and
$\omb>1$ than for $\omb<1$.

Finally, like in one dimension,\cite{HoEvvdL03} $\Ek$ remains small but
finite even at very strong coupling, which is a consequence of the fast and
undirected motion of the electron inside the polaron. As pointed out by
Kornilovitch,\cite{Ko97} the kinetic energy in the small polaron regime is
therefore not related to the effective mass of the electron, the latter
exhibiting an exponential decrease as a function of $\lambda$
(Refs.~\onlinecite{Ko98},~\onlinecite{JeWh98},~\onlinecite{KuTrBo02},
and~\onlinecite{RoBrLi99III}).

To address the issue of finite-size effects, we have calculated $\Ek$ for
$\beta t=10$, $\omb=0.1$ and different linear lattice sizes $N=4$\,--\,12
[see Fig.~\ref{fig:Ek_2d_size}(a)]. The choice of $\omb=0.1$ is reasonable
since the large polaron, which exists for $\lambda<\lc$, is most extended for
small phonon frequencies, as discussed in Sec.~\ref{sec:holstein}, so that
finite-size effects can be expected to be largest. To illustrate this point,
we also present results for a larger phonon frequency $\omb=2.0$
[Fig.~\ref{fig:Ek_2d_size}(b)]. In the latter case, the local oscillators can
respond very quickly to the motion of the electron, and the extension of the
phonon cloud or lattice distortion surrounding the electron is much smaller.
Since for $\omb>1$ the transition to a small polaron happens at larger
$\lambda$, the polaron will be larger for intermediate values
$1\lesssim\lambda\lesssim2$ compared to the adiabatic regime, in which
$\lc=1$. In total, we therefore expect to have smaller finite-size effects
for $\omb=2.0$ than for $\omb=0.1$ as long as we are in the weak-coupling
regime $\lambda\lesssim1$, while the opposite should be true for
$1\lesssim\lambda\lesssim2$. For $\lambda>2$, a highly immobile polaron state
exists in both cases and results should therefore be virtually independent of
$N$. All this is well confirmed by the results shown in
Fig.~\ref{fig:Ek_2d_size}.

Figure~\ref{fig:Ek_2d_size} also reveals that the results begin to saturate
for $N\geq8$, as pointed out previously by Kornilovitch.\cite{Ko97} However,
in contrast to the one-dimensional case,\cite{HoEvvdL03} where we performed
simulations for $N$ as large as 32, we find a nonnegligible dependence on $N$
up to the largest system size ($N=12$). This is better illustrated in
Fig.~\ref{fig:Ek_finite_size}, in which we show $\Ek$ as a function of $1/N$,
and for several values of $\lambda$. To allow for a better representation,
some curves have been shifted, as indicated in the legend. From
Fig.~\ref{fig:Ek_finite_size}, we see that $\Ek$ changes very little for
$N>4$ for the case of strong coupling $\lambda=2$ and, in fact, remains
constant within the error bars, similar to the 1D results in I. For smaller
values of $\lambda$, no such saturation is found on the scale of
Fig.~\ref{fig:Ek_finite_size}, and the behavior of the kinetic energy for
large $N$ is almost linear when plotted as a function of $1/N$. We used a
linear fit of the data for $N=8$, 10, and 12 to obtain an approximation to
the thermodynamic limit. The results are shown in Fig.~\ref{fig:Ek_2d_size}.
Obviously, $\Ek$ for $N=\infty$ has decreased noticeably for small values of
$\lambda$ (including $\lambda=0$), while it remains almost unchanged in the
small polaron regime.  The decrease of $\Ek$ for $\lambda<\lc$ can easily be
understood if we consider the fact that our method works at a finite
temperature $1/\beta$. As a consequence, for very small $N$, the energy gap
between the ground state with $\bm{k}=0$ and the first excited state with
$\bm{k}\neq0$ is larger than the thermal energy $(\beta t)^{-1}$. With
increasing system size, thermal population of excited states becomes
possible. For $N$ large enough ($N\approx20$ for $\beta t=10$ and
$\lambda=0$), results converge to those for $N=\infty$ and, in fact, the
extrapolated data for $\lambda=0$, shown in Fig.~\ref{fig:Ek_finite_size},
agree well with the results for a free electron on an infinite lattice. We
ascribe the smallness of this finite-temperature effect in the
strong-coupling regime above $\lc$ to the very small width of the polaron
band (see, \eg, Ref.~\onlinecite{HoAivdL03}). Consequently, the low-energy
coherent states with different $\bm{k}$ have very similar energies.

Figure~\ref{fig:Ek_2d_size}(a) also shows that for small phonon frequencies,
the influence of the lattice size is largest near $\lc$. The cross over is
sharper for small systems than for the case of large $N$. A similar behavior
has been found by Marsiglio for the one-dimensional model.\cite{Marsiglio95}

Finally, as pointed out by Kornilovitch, \cite{Ko97} despite the good
convergence of the results for quantities such as the kinetic energy even for
small $N\simeq4$, other observables, such as the effective mass, rely on the
knowledge of the energy of states with infinitely small momenta. Therefore,
they will show a more pronounced dependence on $N$ when calculated on small
clusters accessible with, \eg, ED.

We also investigated the effect of temperature on our results, again for
$\omb=0.1$ and $\omb=2.0$. The results, shown in Fig.~\ref{fig:Ek_2d_beta},
indicate that $\Ek$ is more affected by the finite temperature of the
simulation in the adiabatic case $\omb=0.1$ [Fig.~\ref{fig:Ek_2d_beta}(a)].
This is a consequence of the fact that calculations at finite temperatures
only give ground-state-like results when $\beta\om\gg1$. Clearly, this
condition is much more difficult to meet for $\omb=0.1$, and requires larger
values $\beta t>10$.

The changes of $\Ek$ with temperature result from an interplay of several
effects. For $\lambda=0$, the kinetic energy approaches its full
noninteracting value of $-2t\rD$ (\ie, $\Ek=1$) as $T\rightarrow0$. At finite
$T$, however, states with nonzero total quasimomentum $\bm{k}$ will
contribute and thereby lead to a decrease of $\Ek$. As discussed above, this
effect of temperature on $\Ek$ is expected to decrease with increasing
$\lambda$, and to be extremely small in the strong-coupling regime. In the
adiabatic case [Fig.~\ref{fig:Ek_2d_beta}(a)], the transition at $\lc=1$ is
smeared out at high temperatures. For both $\omb=0.1$ and $\omb=2.0$, a
qualitative change in behavior occurs near $\lc$. $\Ek$ decreases with
increasing temperature for $\lambda<\lc$, whereas the opposite is true for
$\lambda>\lc$.  The behavior above $\lc$ can be understood by considering the
electronic hopping amplitude, given by the overlap of the wave functions of a
displaced and an undisplaced oscillator at neighboring sites.  While the
latter is exponentially reduced with increasing electron-phonon coupling at
$T=0$ (see above), it increases with temperature since the oscillators can
occupy excited states, corresponding to wave functions which are more spread
out than the ground state. Moreover, the thermal energy allows the electron
to overcome the potential barrier more easily.  This thermally activated
hopping---which shows up more clearly for small phonon frequencies since, in
this case, phonon excitations require less energy---has been studied already
a long time ago by Holstein.\cite{Ho59a}

It is interesting to notice that the dependence of $\Ek$ on temperature, as
shown in Fig.~\ref{fig:Ek_2d_beta_scaling}, is almost linear at low
temperatures and for $\omb=2.0$ [see Fig.~\ref{fig:Ek_2d_beta_scaling}(b)],
while in the adiabatic case [Fig.~\ref{fig:Ek_2d_beta_scaling}(a)] this is
only true for strong electron-phonon coupling. However, as pointed out
before, for $\omb=0.1$, an inverse temperature $\beta t=10$ is not sufficient
to obtain well-converged results. Therefore, a linear dependence, as for
$\omb=2.0$, may still be found at lower temperatures.  Such calculations,
with the accuracy of the results presented here, would be more
time-consuming. The data for $N=12$, $\beta=10$, $\dtau=0.05$, and
$\lambda=1.0$, \ie, in the cross over regime where statistical errors are
largest (about 0.5\%), required about 28 days of CPU time on an Intel Xeon
2600 MHz computer. However, as has been mentioned previously, the numerical
effort may be reduced by a factor of order $N^\rD$ by linearizing the
exponential of the hopping matrix, denoted as $\kappa$ (see
Sec.~\ref{sec:numerics}.)  Similar to the finite-size scaling performed
above, we also linearly extrapolated the data for $\beta t=10$ and 15 to the
zero-temperature limit $\beta t=\infty$. The results for $\omb=2.0$, for
which such a linear scaling is reasonable, are shown in
Fig.~\ref{fig:Ek_2d_beta}(b). While the general trend agrees well with our
expectations based on the finite-temperature data shown, the scaling
procedure clearly overestimates the temperature effects, thereby leading to
spurious values $\Ek>1$ at $\lambda=0$. However, this can easily be
understood keeping in mind that results saturate at low-enough values of
$\beta t$, so that the linear extrapolation used here becomes insufficient.

Finally, we would like to mention that the results of de Raedt and
Lagendijk\cite{dRLa82,dRLa83,dRLa84} were also given for $\beta t=5$ (and
$\omb=1$), but the small number of Trotter slices ($L=32$ in their case)
gives rise to relatively large systematic errors.\cite{Ko97} This is not the
case for the results of Ref.~\onlinecite{Ko97}, in which the same
extrapolation to $\dtau=0$ was employed as here. Finally, the method of
Kornilovitch\cite{Ko97,Ko98} is free of such errors, but only permits one to
calculate ground-state properties for a restricted range of $\omb$ and
$\lambda$.

\subsubsection{Three dimensions}

\begin{figure}
  \includegraphics[width=0.45\textwidth]{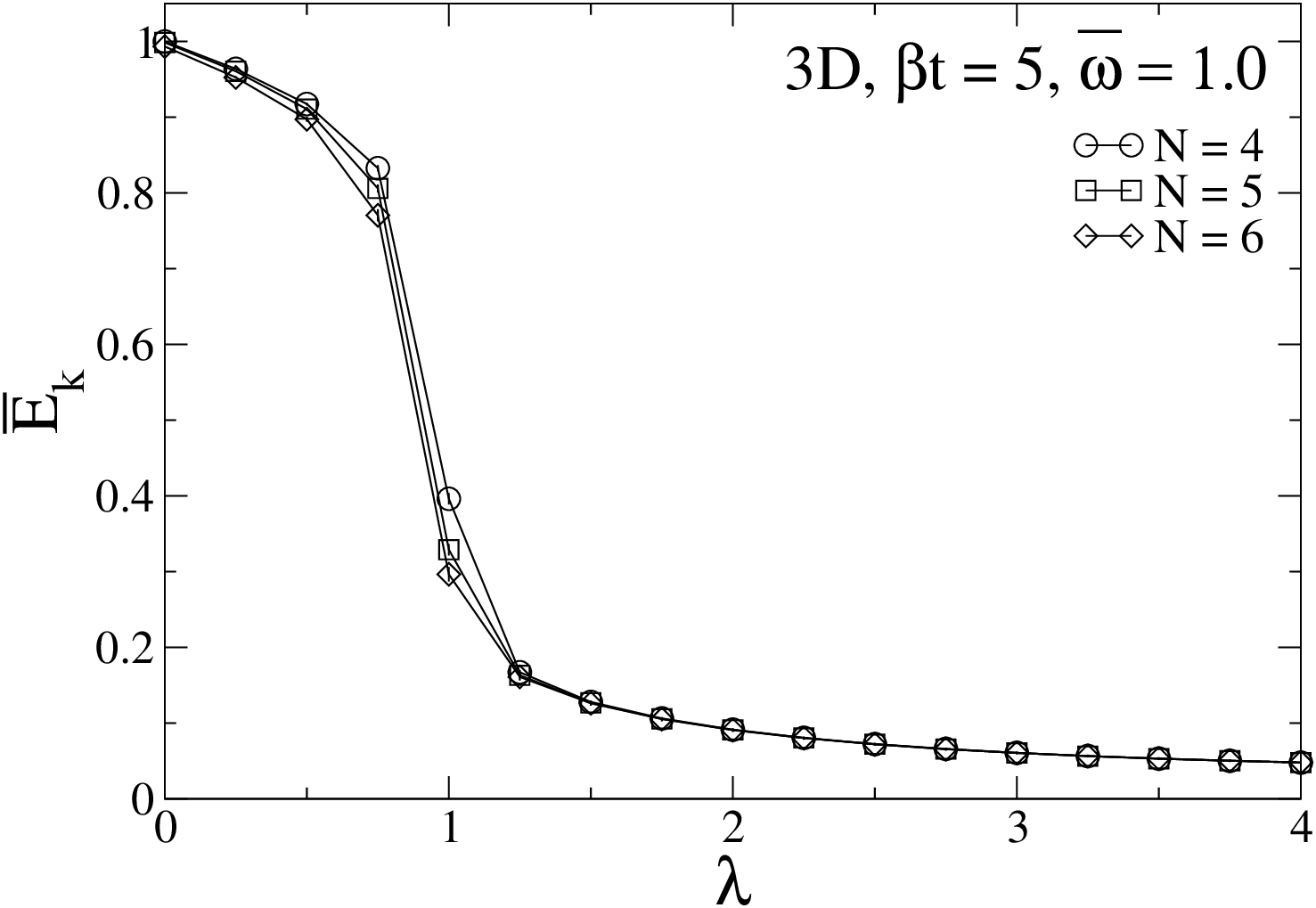}
  \caption{\label{fig:Ek_3d_various_N}
  Normalized kinetic energy $\Ek$ as a function of electron-phonon coupling
  $\lambda$ for different linear dimensions $N$ of the lattice.}
\end{figure}
\begin{figure}
  \includegraphics[width=0.45\textwidth]{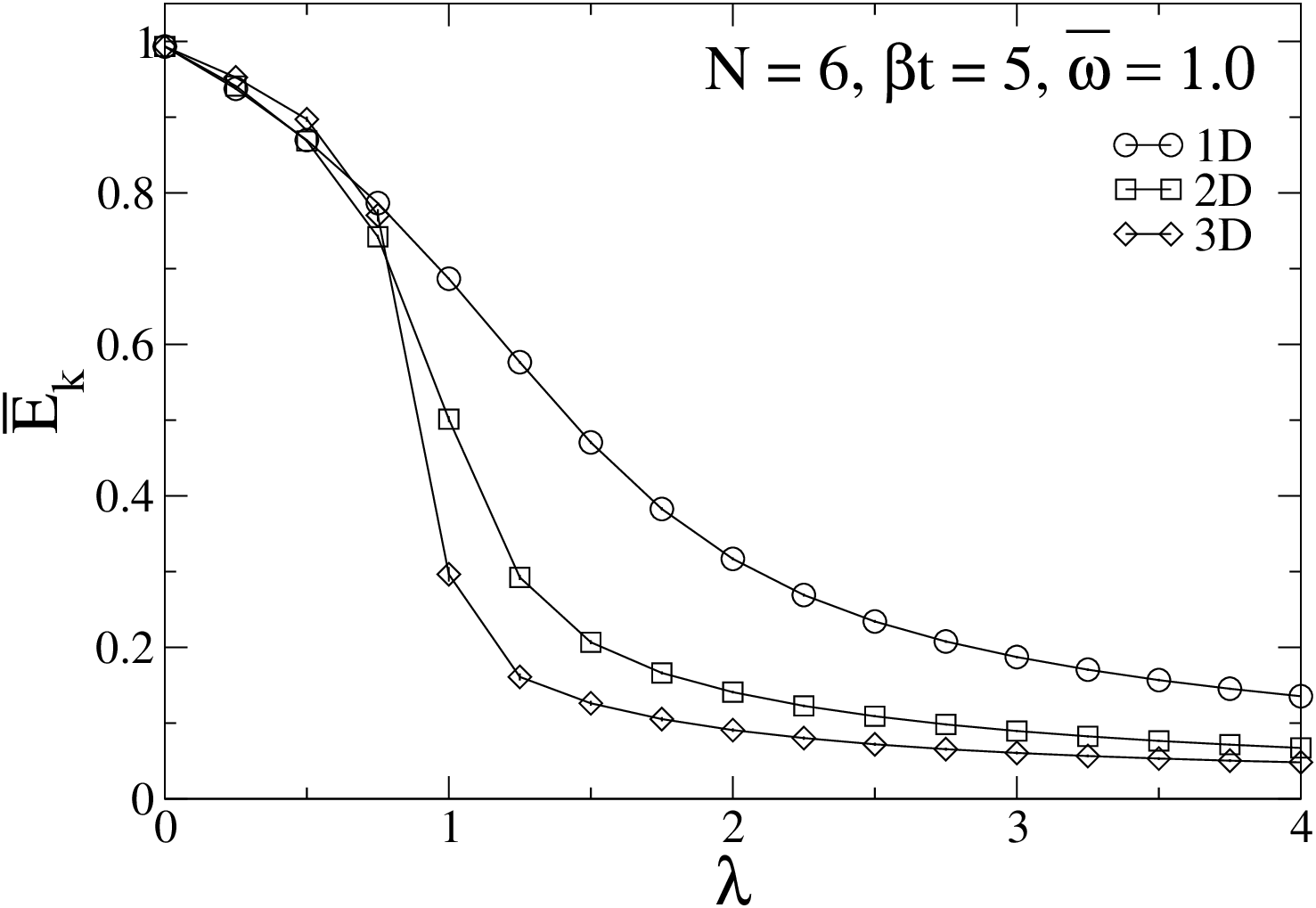}
  \caption{\label{fig:Ek_dimensionality}
    Normalized kinetic energy $\Ek$ as a function of electron-phonon coupling
    $\lambda$ for different dimensions D of the lattice.}
\end{figure}
In contrast to the two-dimensional case discussed above, less work has been
done in three
dimensions.\cite{dRLa82,dRLa83,dRLa84,Ko98,Ko99,RoBrLi99,RoBrLi99III,KuTrBo02}
In fact, we are only aware of one calculation of the kinetic energy, which is
by de Raedt and Lagendijk (DRL).\cite{dRLa83} To compare with their work, we
chose the same values for the phonon frequency $\omb=1.0$ and temperature
$\beta t=5$.  As pointed out in Sec.~\ref{sec:QMC}, the numerical effort for
calculations with our method, which is proportional to $N^{3\rD}$ for the
algorithm in the form used here but could be reduced to $N^{2\rD}$, restricts
us to smaller systems than those considered by
DRL.\cite{dRLa82,dRLa83,dRLa84} For simplicity, we have therefore limited
ourselves to a maximum of $N=6$, for which results can easily be obtained
within a reasonable amount of computer time, while the data presented in
Refs.~\onlinecite{dRLa82,dRLa83,dRLa84} is for $N=32$. To be more specific,
our calculations for one value of $\lambda$, for $N=6$ and $\dtau=0.05$, took
about 10 h on an Intel Xeon 2600 MHz computer.  Due to the relatively small
system size in our work, it is important to study to what extent the results
are converged with respect to $N$. To this end, in
Fig.~\ref{fig:Ek_3d_various_N}, we present $\Ek$ as a function of $\lambda$
for $N=4,5$, and 6. Surprisingly, the results are already satisfactorily
converged, keeping in mind the rather small values of $N$. There is a maximal
change of less than 20\% in the transition region at $\lambda=1$, while $\Ek$
remains almost constant for small and large $\lambda$, as the linear size
increases from $N=4$ to $N=6$. Thus, increasing $N$ further will not change
the results qualitatively, although the finite temperature of our simulations
will manifest itself in a way similar to the two-dimensional case. Our
findings agree well with the results of DRL.\cite{dRLa83} The main difference
is that for weak coupling, our results are closer to the zero-temperature
values (\eg, $\Ek=1$ at $\lambda=0$). The reason for this
discrepancy---despite the fact that we have used the same temperature---is
the smaller lattice size in our calculations. In contrast to the
two-dimensional case considered above, we have not performed a scaling to
$N=\infty$ in 3D, since the clusters under consideration are too small to
reveal a systematic power law dependence as a function of $1/N$.

Finally, we wish to investigate the effect of dimensionality on the
small-polaron cross over. Therefore, we compare $\Ek$ in one, two and three
dimensions using $\beta t=5$, $N=6$, and $\omb=1.0$. The dependence on D,
which is shown in Fig.~\ref{fig:Ek_dimensionality}, is in perfect agreement
with previous work. The transition from a mobile large polaron to a small
polaron, moving in a very narrow but still coherent band, sharpens
considerably with increasing dimension of the system, and while $\Ek$ only
displays a gradual decrease in 1D---without any signs of an abrupt change at
$\lambda=1$---we find a sharp and well-defined transition in three
dimensions.

We conclude this section by comparing the accuracy of our results with the
QMC methods of DRL,\cite{dRLa82,dRLa83,dRLa84,Ko97} and
Kornilovitch.\cite{Ko98,Ko99} As discussed in Sec.~\ref{sec:QMC}, their main
advantage over our method is the fact that they allow one to obtain data
which are essentially free of finite-size effects, in any dimension
$\rD=1$\,--\,3, and with modest computational effort. However, we have seen
above that even in three dimensions, where the limitation of our algorithm is
most noticeable, results are reasonably converged. While the approach of
Refs.~\onlinecite{Ko98} and~\onlinecite{Ko99} is limited to $T=0$, DRL's
method as well as the current approach can, in principle, be used to study
any temperature. Apart from the sign problem discussed in
Sec.~\ref{sec:sign}, the only limitation which occurs is the fact that one
has to increase the number of Trotter slices as $\beta\rightarrow\infty$, so
as to keep the Trotter error smaller than the statistical errors. This
situation can be greatly improved by extrapolating results to $\dtau=0$, as
has been done in this work, while the continuous-time algorithm of
Kornilovitch\cite{Ko98,Ko99} is free of any such discretization errors.  The
accuracy of the results presented here depends on $\omb$, $\beta t$, $N$ and
$\lambda$. Away from $\lambda\approx1$, error bars are use usually smaller
than the line width, corresponding to relative errors of less than 0.5\%.
This is comparable to the accuracy of the results given by
Kornilovitch\cite{Ko98} and significantly more accurate than the original
results by DRL.\cite{dRLa82,dRLa83}

\section{\label{sec:summary}Conclusions}

Extending the work of Ref.~\onlinecite{HoEvvdL03} to the case of the two- and
three dimensional Holstein model with one electron, we have shown that our
QMC approach allows accurate calculations with modest computational effort
for a large range of parameters. In particular, the minus-sign problem, due
to the Lang-Firsov transformation, has been found to diminish quickly with
increasing system size, and not to have a significant effect on simulations.
We have presented a detailed study of the small-polaron cross over and its
dependence on the parameters of the system. In particular, we have focused on
finite-size and finite-temperature effects, which have not been investigated
systematically before.

As discussed above, our approach is not as fast as other QMC methods for the
Holstein polaron,\cite{dRLa82,dRLa83,dRLa84,Ko97,Ko98,Ko99} the main
limitation being the restriction to smaller but still reasonably large
lattices. This difference in performance is acceptable keeping in mind that
it can be extended to the many-electron case (see below), in contrast to the
world-line methods of Refs.~\onlinecite{dRLa82,dRLa83,dRLa84,Ko97,Ko98,Ko99},
which work best for the case of a single electron or two electrons of
opposite spin, and face a severe sign problem in more than one dimension,
similar to other world-line methods.\cite{wvl1992} Finally, despite the sign
problem, we can perform accurate simulations for virtually all physically
interesting values of $\omb$ and $\lambda$, in contrast to the method of
Refs.~\onlinecite{Ko98} and~\onlinecite{Ko99}.

Motivated by the promising results of this work and
Ref.~\onlinecite{HoEvvdL03}, our next objective will be a generalization to
the many-electron case. To this end, it is important to notice the striking
similarity of the QMC method presented here with the determinant method first
introduced by Blankenbecler \etal,\cite{BlScSu81} which has been successfully
used to study superconductivity and charge-density-wave formation in the
Holstein model.\cite{BlScSu81,ScSu81,LeSu90,LeSu91,NiGuScFo93} In fact, the
main difference between our one-electron algorithm and the corresponding
grand-canonical method is the calculation of the fermionic weight, which in
case of the latter is given by the determinant of $\hat{1} + \Omega$. Here
$\hat{1}$ denotes the unity operator, and $\Omega$ is the corresponding
generalization of Eq.~(\ref{eq:wf}) to many electrons.  Consequently, the
numerical effort is expected to be similar to the simulations with one
electron. However, an important open question concerns the dependence of the
sign problem on the number of electrons. This work is currently in progress.

\begin{acknowledgments}
  
  This work has been supported by the Austrian Science Fund (FWF), project
  No.~P15834. One of us (M.H.) is grateful to the Austrian Academy of
  Sciences for financial support. The majority of the results presented here
  have been obtained using the Gigabit-Cluster at Graz University of
  Technology.

\end{acknowledgments}
  


\end{document}